\documentclass[%
reprint,
longbibliography,
amsmath,amssymb,
aps,
]{revtex4-1}

\usepackage{graphicx}
\usepackage{dcolumn}
\usepackage{bm}
\usepackage{xcolor}

\newcommand{\E}{\varepsilon}

\newcommand{\B}{\beta}

\begin{document}
	
	\preprint{APS/123-QED}
	
	\title{Constructing near-field and far-field with reactive metagratings:\\
	study on degrees of freedom}

		\author{Vladislav Popov}
	\email{uladzislau.papou@centralesupelec.fr}
	\affiliation{%
		SONDRA, CentraleSup\'elec, Universit\'e Paris-Saclay, F-91190, Gif-sur-Yvette, France
	}%
	\author{Fabrice Boust}%
	\email{fabrice.boust@onera.fr}
	\affiliation{%
		SONDRA, CentraleSup\'elec, Universit\'e Paris-Saclay,
		F-91190, Gif-sur-Yvette, France
	}
	\affiliation{%
		DEMR, ONERA, Universit\'e Paris-Saclay, F-91123, Palaiseau, France
	}%
	\author{Shah Nawaz Burokur}%
	\email{sburokur@parisnanterre.fr}
	\affiliation{%
		LEME, UPL, Univ Paris Nanterre, F92410, Ville d'Avray, France
		}%

\begin{abstract}
In this paper, we report that metamaterials-inspired one-dimensional gratings (or metagratings) can be used  to control  nonpropagating diffraction orders  as well as propagating ones. 
By accurately engineering the near-field it becomes possible to satisfy power conservation conditions and achieve perfect control over all propagating diffraction orders with passive and lossless metagratings. 
We show that each propagating diffraction order requires two degrees of freedom represented by passive and lossless loaded thin ``wires''. It  provides a solution to the old problem of power management between diffraction orders created by a grating. The developed theory is verified by both 3D full-wave numerical simulations and experimental measurements, and can be readily applied to the design of wavefront manipulation devices over the entire electromagnetic spectrum as well as in different fields of physics.
\end{abstract}

\pacs{42.25.Bs, 78.67.Pt, 81.05.Xj}

\maketitle

\section{introduction}

Back at the beginning of the 20$^\textup{th}$  century, the problem of intensity distribution among different diffraction orders produced by a diffraction grating was one of the most important in optics~\cite{Wood1910}. 
Since then, a particular class of grating maximizing the intensity in a given diffraction order referred to as blazed gratings was studied in detail~\cite{Wood1910,Rowland1893,stamm1946energy,Breidne1979} and perfect blazing was demonstrated 
in nonspecular direction when only two orders propagate~\cite{Hessel1975,Breidne1981}.
In the context of antenna applications, highly efficient reflection and transmission at small  diffraction angles was achieved by means of classical reflect- and transmit-arrays ~\cite{Pozar1996,huang2007reflectarray,Pozar2007}.

Amazing possibilities in manipulation of electromagnetic fields with engineered dense distributions of scatterers (metamaterials) have been demonstrated in the last two decades  \cite{Jacob2016,Glybovski2016,C7TC03384B,Tong2018}.
Extensive research in the area of metasurfaces, thin two-dimensional metamaterials, established a rigorous theoretical approach to arbitrary control  reflection and refraction of an incident plane-wave ~\cite{Capasso_GeneralizedReflectionLaw,Pfeiffer2013,Asadchy2016_SpatiallyDispMS,Asadchy2017MultiChannel}. 
In what follows we discuss examples of the perfect control (without spurious scattering) over the reflection/transmission that were demonstrated by means of a rigorous theory.
Thus, perfect refraction in the first diffraction order and beam splitting in transmission with equal excitation of -1$^\textup{st}$ and 1$^\textup{st}$ diffraction orders by means of passive and lossless bianisotropic metasurfaces was presented in~\cite{Epstein2016_fieldTrans_OBMS,Epstein2018_exp_anrefr,8259235} and \cite{Epstein2016_AuxiliryFields}, respectively. 
In order to perform perfect nonspecular reflection with passive and lossless metasurfaces, auxiliary surface waves have to be additionally excited ~\cite{Epstein2016_AuxiliryFields,Tretyakov_2017_auxilirySW,Tretyakov2017_perfectAR,Kwon2018}. 
Although it seems possible to design such metasurfaces the design procedure is still not well established~\cite{Tretyakov2017_perfectAR,Kwon2018}. 
Huygens' metasurfaces  having equivalent electric and magnetic responses allow one to  efficiently control diffraction from microwave~\cite{Pfeiffer2013} to optical frequencies~\cite{PhysRevLett.110.203903}   under the  conditions of local normal power flow conservation and conjugate impedance matching~\cite{Epstein2014_HMS_diffraction}.

In Ref.~\cite{Alu2017_metagrating},  \textit{Ra’di et al.} have recently introduced the concept of metagratings  which are an evolution of conventional one-dimensional (1D) diffraction gratings. The prefix ``meta'' implies that the grating is constructed from meta-atoms whose scattering properties can be judiciously engineered.
Traditionally, in 1D gratings there is a profile modulation in one direction and a translational symmetry in the other. In metagratings, the translation invariant direction is engineered at a scale that is small compared to the wavelength such that it becomes possible to define an averaged macroscopic quantity like an impedance density~\cite{Epstein2017}. The possibility to engineer the impedance density and an accurate analytical model allows one to overcome the limitations of metasurfaces. For instance, in Refs.~\cite{Alu2017_metagrating,Epstein2017,Epstein2018} the authors, by means of theory and full-wave simulations, demonstrated the possibility of perfect nonspecular reflection and beam splitting in reflection with a metagrating composed of only a single unit cell per period.
In order to realize perfect refraction in the  1$^\textup{st}$ diffraction order three unit cells per period are required, as was shown analytically in~\cite{epstein2018anrefr}. 
Lately, experimental verification of perfect reflection in the -1$^\textup{st}$ diffraction order has been reported by \textit{Rabinovich et al.} in Ref.~\cite{rabinovich2018experimental}.
In Refs.~\cite{Eleftheriades_2018,Eleftheriades_2018_1} the authors numerically and experimentally demonstrated  efficient  broadband nonspecular reflection with a $2$-cell periodic structure capable of controlling  two propagating diffraction orders.

The way towards  control over arbitrary number of propagating diffraction orders by means of many unit cells based metagratings was outlined in Ref.~\cite{Popov2018} for a reflection configuration. Moreover, it was shown that when the number of degrees of freedom is equal to the number of propagating diffraction orders, perfect total control is possible only in the case when engineered active and lossy responses are available. Otherwise, there are scattering losses. 
In this paper, we report that metagratings can be used  to control  nonpropagating diffraction orders  as well as propagating ones. 
By accurately engineering the near-field it becomes possible to satisfy power conservation conditions and achieve perfect control over all propagating diffraction orders with passive and lossless metagratings. 
In what follows, we study  theoretically and validate experimentally the number of degrees of freedom required by each propagating diffraction order thus  providing a solution to the old problem of power management between diffraction orders created by a grating.

\begin{figure}[tb]
\includegraphics[width=0.8\linewidth]{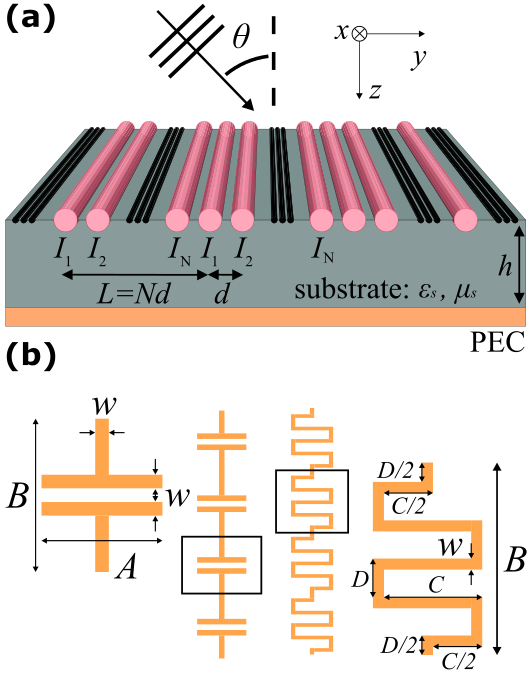}
\caption{\label{fig:1} Schematics of the considered physical system and elements of its practical implementation. \textbf{(a)} System under consideration: a periodic array of thin wires [large cylinders] placed on a PEC-backed dielectric substrate having relative permittivity $\E_s$, permeability $\mu_s$ and thickness $h$. The array is excited by a TE-polarized plane-wave incident at an angle $\theta$. \textbf{(b)} Schematics of implementation at microwave frequencies of capacitively (left) and inductively (right) loaded PEC strips.}
\end{figure}

\section{perfect  control of diffraction: \\two reactive elements per an order}

Theoretically, a metagrating is described as a one-dimensional periodic  array of polarization line currents which are excited in thin loaded ``wires'' by a TE-polarized  plane-wave   incident at an angle $\theta$ and having the electric field along the wires. We consider a reflective-type metagrating when the  wires are placed on top of a perfect electric conductor (PEC)-backed dielectric substrate. Schematics of the system under consideration is depicted in Fig.~\ref{fig:1} (a). 
A grounded substrate should be carefully chosen in order to provide efficient excitation of line currents [i.e. $h\approx \lambda/(4\sqrt{\E_s\mu_s-\sin(\theta)^2}$)] and avoid excitation of waveguide modes~\cite{Popov2018}.


Since the illuminated structure is periodic the  wave reflected outside the substrate [$z<-h$] can be represented as a superposition of plane-waves
$
\sum_{m=-\infty}^{+\infty}A_m^{TE}e^{-j\xi_my+j\B_mz}
$.
The plane-waves have the tangential and normal components of wave vector equal to $\xi_m=k\sin[\theta]+2\pi m/L$ and $\B_m=\sqrt{k^2-\xi_m^2}$, respectively, with $k$ being the wavenumber outside the substrate. A simple model of metagratings allows one to find the amplitudes $A_m^{TE}$ analytically (see details in Appendix~\ref{app:a})
\begin{eqnarray}\label{eq:Am}
&&A_m^{TE}=-\frac{k\eta}{2L}\frac{
(1+R_m^{TE})e^{j\B_mh}}{\B_m}\sum_{q=1}^{N}I_qe^{j\xi_m (q-1)d}\nonumber\\
&&+\delta_{m0}R_0^{TE}e^{2j\B_0h}
\end{eqnarray}
where $\eta=\sqrt{\mu/\varepsilon}$ is the characteristic impedance outside the substrate, $\delta_{m0}$ represents the reflection of the incident wave from the grounded substrate and $R_m^{TE}$ is the Fresnel's reflection coefficient.
Equation ~\eqref{eq:Am} reveals that each of $N$ line currents in a supercell  contributes to the reflected plane-waves through the discrete Fourier transformation of the sequence $I_q$. Although there is an infinity of reflected plane-waves, only a finite number $M=r+l+1$ of them is scattered in the far-field determining the diffraction pattern. $r$ and $l$ are the  largest integers such that $\B_{r}>0$ and $\B_{-l}>0$. Currents $I_q$  represent  degrees of freedom that can be harnessed to control the amplitudes of the reflected fields as seen from equation~\eqref{eq:Am}.

\begin{figure}[tb]
\includegraphics[width=0.99\linewidth]{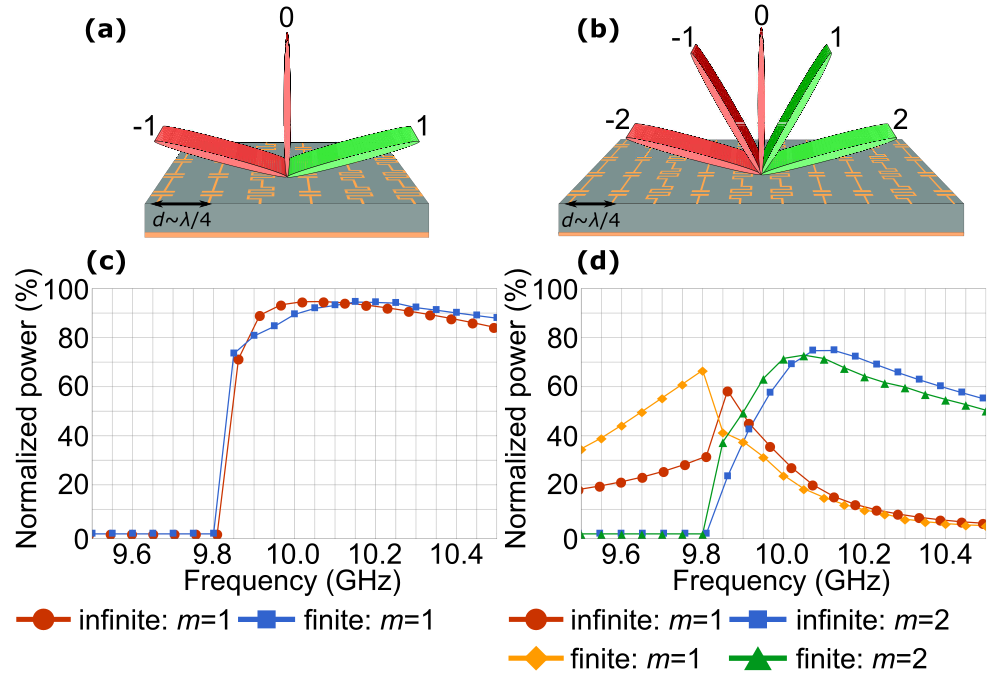}
\caption{\label{fig:2} 
Power management between propagating diffraction orders by  the considered metagratings  with six and eight unit cells per period: schematics (top row) and simulation data (bottom row).
Result for infinite and finite size metagratings are presented.
Figures in the top row  depict excited (green lobes) and canceled (red lobes) propagating diffraction orders corresponding to the plots in the bottom row showing the 3D full-wave simulated frequency responses of the metagratings (i.e. part of total power scattered in a given diffraction order versus frequency).
\textbf{(a)}, \textbf{(c)} Example of nonspecular reflection at an angle of $80^o$ by means of a metagrating with $N=6$ unit cells per period. The finite size metagrating has  $16$ supercells.
\textbf{(b)}, \textbf{(d)} Example when out of five plane-waves reflected in the far-field, only the first ($1/3$ of total power) and second ($2/3$ of total power) propagating diffraction orders are excited  with a metagrating having  $N=10$ unit cells in a period. The finite size metagrating has  $8$ supercells.
In both examples, normal incidence is assumed.
}
\end{figure}

Each polarization line current is excited in a thin wire  characterized by  its input-impedance  $Z_{in}$ and load-impedance  $Z_q$   densities. Necessary  currents $I_q$ can be obtained by loading wires with suitable load-impedance densities $Z_q$ which are found from the following equation \begin{equation}\label{eq:Z}
Z_qI_q=E_{q}^{(exc)}-Z_{in}I_q-\sum_{p=1}^N Z_{qp}^{(m)}I_p.
\end{equation}
The right-hand side of equation~\eqref{eq:Z} represents the total electric field at the location of the $q^\textup{th}$ wire, $E_{q}^{(exc)}$ represents the excitation field (incident wave plus the wave reflected from the grounded substrate), $Z_{qp}^{(m)}$ are the mutual-impedance densities which account for the interaction between the wires and between the wires and the grounded substrate. 
The details on the derivation of Eqs.~\eqref{eq:Am} and \eqref{eq:Z} as well as  the explicit expressions of the impedance densities can be found in Ref.~\cite{Popov2018}. For sake of the reader's convenience we place main parts of the derivations in Appendix~\ref{app:a}.

Total control of the diffraction pattern is possible by $M$ line currents per supercell.
However, we are particularly interested in purely reactive solutions of equation~\eqref{eq:Z}, since in practice it can be challenging to engineer active/lossy response of the load.
Thus,  the currents $I_q$ should also satisfy the conditions of passivity and absence of loss
\begin{eqnarray}\label{eq:reactive}
\Re\left[\left(E_{q}^{(exc)}-\sum_{p=1}^N Z_{qp}^{(m)}I_p\right)I_q^*\right]=\Re[Z_{in}]|I_q|^2,
\end{eqnarray}
where the  asterisk symbol stands for the complex conjugate.
Equation~\eqref{eq:reactive} represents a set of $N$ quadratic algebraic equations with real and imaginary parts of currents being the variables and simply means that the $q^\textup{th}$ current  radiates all the power spent on its excitation. 
Additional $M$ (complex-valued) line currents are required to satisfy equation~\eqref{eq:reactive}. 
Thus,  $N=2M$ line currents per supercell are necessary for establishing  arbitrary diffraction patterns exactly.
Although there can be many line currents in a period, the distance between them is of the order of $\lambda/4$ ($\lambda$ is the operating wavelength), which does not allow one to perform homogenization and introduce surface impedance.

From the physical point of view, the additional $M$ currents are used to set the amplitudes $A^{TE}_m$ of the surface waves (or nonpropagating diffraction orders, $m>r$ and $m<-l$) which would ensure equation~\eqref{eq:reactive}. For a better understanding, let us consider an example of a perfect reflection in the 1$^\textup{st}$ diffraction order of a plane-wave at normal incidence ($r=l=1$). In this case, one has to cancel two propagating diffraction orders since there are three plane-waves reflected in the far-field and thus, the necessary number $N$ of line currents per period  is $6$. First of all, one sets the amplitudes of the plane-waves in the far-field as $A_{-1}^{TE}=0$, $A_{0}^{TE}=0$ and  $A_1^{TE}=e^{j\phi_1}$, where $\phi_1$ is the phase of the anomalously reflected wave. Then the line currents $I_q$ ($q=1,2,...,6$) found from equation~\eqref{eq:Am} ($m=-3,-2,...,2$) are substituted into equation~\eqref{eq:reactive}. The unknown (complex) amplitudes $A_{-3}^{TE}$, $A_{-2}^{TE}$ and $A_{2}^{TE}$ of the surface waves are found by solving equation~\eqref{eq:reactive}, which automatically ensures the passive and lossless load-impedance densities $Z_q$ calculated afterwards from equation~\eqref{eq:Z}.

\begin{figure*}[tb]
\includegraphics[width=0.99\linewidth]{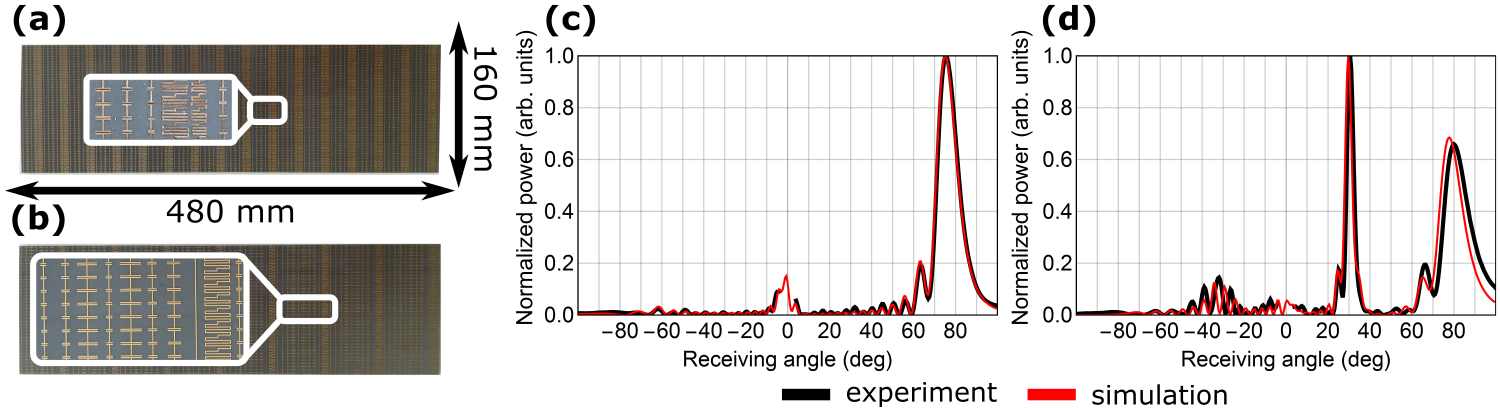}
\caption{\label{fig:3} Fabricated samples and comparison of the simulation and experimental data. \textbf{(a)}, \textbf{(b)} Photographies of the samples performing (a)  nonspecular reflection at $80$ degrees ($N=6$) and (b)  splitting into two plane-waves propagating at $30$ and $80$ degrees ($N=10$). \textbf{(c)}, \textbf{(d)} Experimentally measured and numerically simulated scattering patterns: (c) nonspecular  reflection at $10.1$ GHz (main beam has $93$\% of total power), (d) unequal splitting into two plane-waves at $9.95$ GHz (there are $31.5$\% of power in the $1^\textup{st}$ order and $63.5$\% in the second one). }
\end{figure*}

\section{Design, simulation and experiment}

Once the necessary  load-impedance densities are  known, one has to come up with a practical implementation of the loads. In a general case, capacitive and inductive loads are required for such design implementation. As a proof of concept we demonstrate the design procedure for metagratings operating at microwave frequencies near $10$ GHz. Thin metallic wires are realized as PEC strips having the input-impedance density $Z_{in}=k\eta H_0^{(2)}[kw/4]/4$ with $H_0^{(2)}$  being the Hankel function of the second kind and $w$ being the width of strips. Capacitive and inductive responses can be achieved with the printed microstrip capacitors and inductors schematically shown in  Fig.~\ref{fig:1} (b).
Load-impedance density $Z_c$ of the printed capacitors can be approximately calculated by means of analytical formulas for the grid impedance of a PEC strips capacitive grid~\cite{tretyakov2003analytical,Tretyakov_patches_2008,Tretyakov_MetaAnalyt_2018}
\begin{equation}\label{eq:Zc}
Z_c=-j\kappa_c\frac{\eta_{eff}}{2A\alpha},\quad \alpha=\frac{k_{eff}B}{\pi}\ln\left[\frac{1}{\sin[\frac{\pi w}{2B}]}\right],
\end{equation}
where $A$ is the arms' length, $\eta_{eff}=\eta/\sqrt{\E_{eff}}$, $k_{eff}=k\sqrt{\E_{eff}}$, $\E_{eff}=(1+\E_s)/2$, $\alpha$ is the grid parameter and $B$ is the period along the $x$-direction. The formula~\eqref{eq:Zc} was already used in the context of metagratings in, e.g., \cite{Epstein2018} and~\cite{Popov2018}. Since PEC strips act intrinsically as inductors themselves ($\Im[Z_{in}]>0$), the inductive load can be implemented  by modulating the effective length of the strip through a meandering design process~\cite{wang2018reciprocal,Tretyakov_MetaAnalyt_2018}. Then, the inductive load-impedance $Z_i$ density can be estimated as 
\begin{eqnarray}\label{eq:Zi}
&&Z_i=j\frac{1}{\kappa_i}\frac{l_{eff}\Im[Z_{in}]}{ B},\quad l_{eff}=C\left(\frac BD-1\right),\nonumber\\
&&\Im[Z_{in}]\approx -\frac{k\eta}{2\pi}\left(\ln\left[\frac{kw}{8}\right]+\gamma\right),
\end{eqnarray}
where $l_{eff}$ is the effective length of the meander, $C$ and $D$ are the parameters of the meander [see Fig.~\ref{fig:1} (b)], and $\gamma\approx 0.5772$ is the Euler constant. Formula~\eqref{eq:Zi} is a rough approximation of the inductive load-impedance since it does not take into account the interaction between the meander strips and capacitive response on the incident wave.
Geometrical parameters $w$, $B$ and $D$ are the same for all unit cells and fixed. Parameters  $A$ and $C$ are found from Eqs.~\eqref{eq:Zc} and \eqref{eq:Zi} for each unit cell accordingly to load-impedance densities calculated beforehand.
The last step of the design procedure is to additionally adjust parameters $A$ and $C$ by performing a parametric sweep with respect to the scaling parameters $\kappa_c$ and $\kappa_i$ which are the same for different unit cells.  In contrast to the design procedure of metasurfaces, here we perform simulations of a whole supercell having $\kappa_c$ and $\kappa_i$ as the only two free parameters. This allows one to account for interaction between unit cells and immediately arrive at the ultimate design. For a more detailed description of the design procedure see Appendix~\ref{app:b}.

\begin{figure*}[tb]
\includegraphics[width=0.99\linewidth]{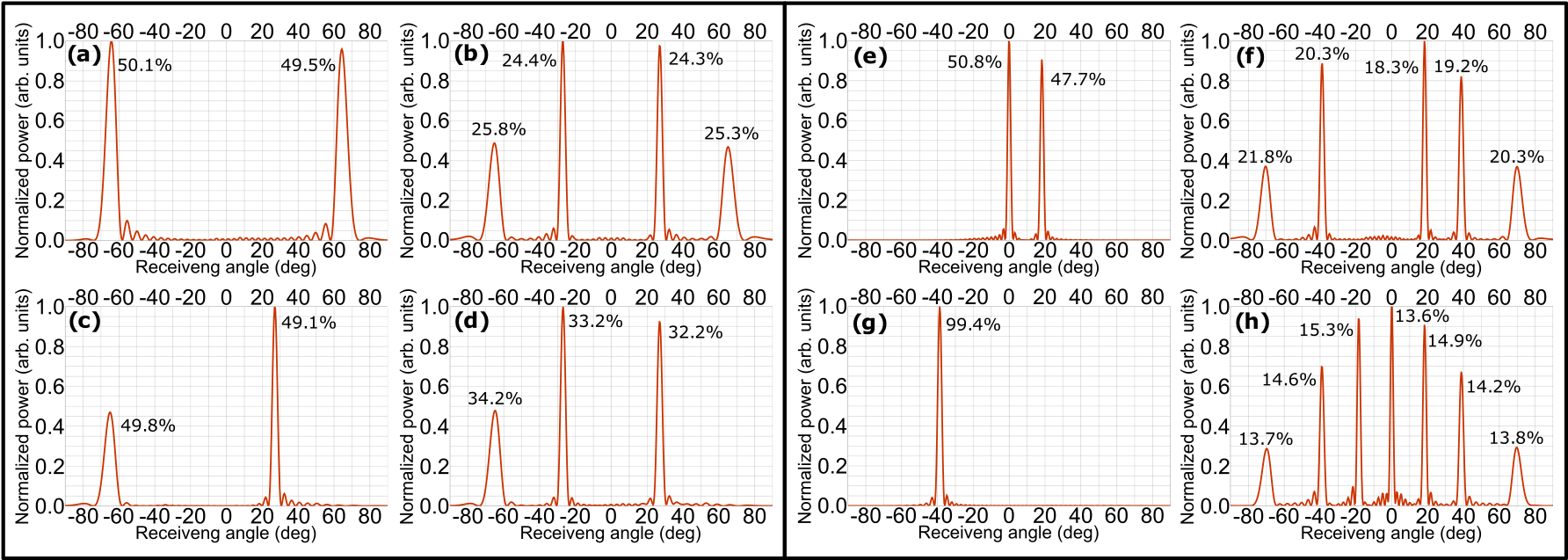}
\caption{\label{fig:4} 
Far-field scattering patterns from finite size metagratings  under normally incident plane-wave obtained by means of  2D full-wave COMSOL simulations. Each finite size metagrating has $8$ supercells. Numbers next to each lobe represent the part of power in a given lobe. All represented examples aim to demonstrate equal distribution of incident power between all excited propagating diffraction patterns (using equation (3) which assumes infinite samples). \textbf{(a)--(d)} Period $L=2\times 30/(\sin(65^\textup o))$, there are five propagating diffraction orders. (a) $-2^\textup{nd}$ and $2^\textup{nd}$ orders are excited. (b) $-2^\textup{nd}$, $-1^\textup{st}$, $1^\textup{st}$ and $2^\textup{nd}$ orders are excited. (c) $-2^\textup{nd}$ and $1^\textup{st}$ orders are excited. (d) $-2^\textup{nd}$, $-1^\textup{st}$ and $1^\textup{st}$ orders are excited. \textbf{(e)--(h)} Period $L=3\times 30/(\sin(70^\textup o))$, there are seven propagating diffraction orders. (e) $0^\textup{th}$ and $1^\textup{st}$ orders are excited. (f) All orders are excited apart from $0^\textup{th}$ and $-1^\textup{st}$. (g) only $-2^\textup{nd}$ order is excited. (h) All propagating orders are excited. }
\end{figure*}

The importance of the near-field control can be demonstrated  by considering a simple example of a nonspecular reflection at extreme angles~\cite{Epstein2016_AuxiliryFields,Tretyakov2017_perfectAR,Kwon2018}. Namely, we consider the reflection of a normally incident plane-wave at the angle of $80$ degrees. In this studied case, there are only three propagating diffraction orders (-1$^\textup{st}$, 0$^\textup{th}$ and 1$^\textup{st}$), as shown by the schematics in Fig.~\ref{fig:2} (a). Thus, for realizing the anomalous reflection one has to cancel scattering in the -1$^\textup{st}$ and 0$^\textup{th}$ diffraction orders, which requires six loaded wires per supercell implemented by passive and lossless elements. 
The second example we consider is the splitting of the normally incident plane-wave into two reflected plane-waves propagating at $30$ (first diffraction order) and $80$ (second diffraction order) degrees. In contrat to commonly demonstrated examples of beam splitting, here the incident wave power is not equally distributed between the excited diffraction orders. Particularly, we design the sample to steer $1/3$ of the total power in the first diffraction order and $2/3$ in the second one. This scenario is schematically depicted in Fig.~\ref{fig:2} (b) where there are five propagating diffraction orders controlled by ten loaded wires in a supercell. Other examples are also provided in Sec.~\ref{sec:4}.

The two metagratings are designed to operate at $10$ GHz and tested in the following three steps. First, by means of 3D full-wave simulations we test the metagratings designs in an infinite array configuration by imposing periodic boundary conditions to a single supercell and by assuming plane-wave illumination. Figures~\ref{fig:2} (c) and (d) demonstrate  the  frequency response of the infinite metagratings. It is seen that the efficiency is above $95$\% in both considered examples at the frequency of operation. The remaining $5$\% power is dissipated as heat in the substrate due to dielectric losses and as spurious scattering due to imperfections of the design. In a second step, 3D full-wave simulations are used to test finite size physical metagratings with a number of supercells corresponding to that used for fabrication of the experimental samples. In order to be able to further compare the results of these simulations to the experimental data, features of the experimental setup have to be taken into account. 
The fabricated samples have been tested in an anechoic chamber dedicated to radar cross section (RCS) bistatic measurements. Transmitting and receiving horn antennas are mounted on a common circular track of $5$ m radius. A photo of the experimental setup is shown in  Appendix.
Physical sizes of the experimental samples are approximately $480$ mm ($y$-direction) by $160$ mm ($x$-direction), as illustrated in Figs.~\ref{fig:3} (a) and (b). Thus the wavefront of the incident wave in the $y$-direction cannot be approximated by a plane-wave. To take this configuration into account, simulations are performed assuming a cylindrical incident wave with periodic boundary conditions applied in the $x$-direction. The scattered fields are calculated on a circle enclosing the metagratings and are then extrapolated to a $5$ m radius with the help of the Chu-Stratton formula~\cite{Chu_Stratton1939,stratton2007electromagnetic}. See Appendices~\ref{app:c} and \ref{app:d} for details on the simulation data processing technique. Figures~\ref{fig:2} (c) and (d) allow one to compare the efficiency of the finite size metagratings with the ideal case of the infinite metagratings.
The discrepancy in Fig.~\ref{fig:2} (d) at low frequencies stems from disappearance of the second orders what, clearly, has an impact on the performance of a finite size metagrating. However, this issue is to be studied yet.
Finally, we compare the simulation results of the finite size metasurfaces with experimental data.
In the current experiment, the transmitter is fixed and the receiver moves with $0.5$ degrees step. The minimum angle value between the transmitter and the receiver for the scanning is $4$ degrees. 
Under this experimental setup configuration, it is not possible to measure specular reflection in the experiment. Therefore, the performance of the fabricated samples can be estimated from the simulation data depicted  in Figs.~\ref{fig:2} (c) and (d). Figures~\ref{fig:3} (c) and (d) compare the measured and simulated scattered patterns,  where a good agreement can be observed. 

\section{Other examples by 2D simulations}\label{sec:4}

So far we have demonstrate only two examples of metagratings for controlling diffraction patterns. However, the developed approach allows one to realize arbitrary diffraction patterns. Figure~\ref{fig:4} demonstrates different configurations of the far-field scattering pattern from metagratings of two different periods. The scattering pattern was obtained with 2D full-wave simulations performed by means of \textup{COMSOL Multiphysics} as described in Appendix~\ref{app:e}. Metagratings in Fig.~\ref{fig:4} were designed to equally split the power of normally incident plane-wave between excited propagating diffraction orders. Numbers next to each lobe represent the part of total power carried by a given beam. The imperfection are only due to the finite size of metagratings in the $y$-direction, i.e. finite number of periods. Indeed, the scattering problem for finite size   objects is more complex than in case of infinite, truly periodic structures. Strictly speaking, the developed theory is valid for finite size metagratings only when an incident wave effectively illuminates a metagrating's area much greater than its period and much less than its whole size. For instance, it is the case for a Gaussian beam with the waist $w_{GB}$ such that $1\ll w_{GB}/L\ll N_s$ ($N_s$ is the total number of supercells).

\section{discussion and conclusion}

\begin{table*}[tb]
\resizebox{\textwidth}{!}{%
\begin{tabular}{|c|c|c|c|c|c|c|c|c|c|c|} 
 \hline
load-impedance density ($\eta/\lambda$) & $Z_1$ & $Z_2$ & $Z_3$ & $Z_4$ & $Z_5$ & $Z_6$ & $Z_7$ & $Z_8$ & $Z_9$ & $Z_{10}$\\ 
 \hline
nonspecular reflection & $-j10.6$ & $-j6.27$ & $-j12.2$ & $j12.5$ & $j22.4$ & $-j15.7$&$-$&$-$&$-$&$-$\\  
 \hline
beam splitting &  $-j9.32$ & $-j6.88$ & $-j2.77$ & $-j8.57$ & $-j2.60$ & $-j6.03$&$-j4.10$&$j0.38$&$j13.0$&$-j8.98$\\ 
 \hline
\hline
geometrical parameters (mm) & $A_1$ & $A_2$ & $A_3$ & $C_4$ & $C_5$ & $A_6$ & $-$ & $-$ & $-$ & $-$\\ 
\hline
nonspecular reflection & $2.0$ & $3.3$ & $1.7$ & $2.9$ & $5.2$ & $1.3$&$-$&$-$&$-$&$-$\\  
 \hline
 \hline
geometrical parameters (mm) & $A_1$ & $A_2$ & $A_3$ & $A_4$ & $A_5$ & $A_6$ & $A_7$ & $C_8$ & $C_9$ & $A_{10}$\\ 
\hline
beam splitting &  $1.7$ & $2.3$ & $5.6$ & $1.8$ & $6.0$ & $2.6$&$3.8$&$0$&$7.0$&$1.7$\\ 
 \hline
 \end{tabular}%
}
\caption{\label{tab:1}Parameters of metagratings presented in the main text. The indexes correspond to the numbered unit cells in Fig.~\ref{fig:1}.}
\label{tab:1}
\end{table*}

The experimental validation results represent extreme examples in the control of diffraction patterns which are challenging or impossible to realize by other means. For instance, in order to perform large angle nonspecular reflection using a scalar reflective metasurface one has to significantly rely on numerical optimization techniques~\cite{Tretyakov2017_perfectAR,Kwon2018}. Otherwise, one has to design a three layer scalar metasurface emulating omega-bianisotropic response or a tensorial reflective metasurface~\cite{Epstein2016_AuxiliryFields,Tretyakov_2017_auxilirySW}. Up to date, neither the bianisotropic nor the tensorial metasurfaces have been validated experimentally or by means of 3D full-wave simulations for nonspecular reflection applications. On the other hand,  metagratings presented in~\cite{Popov2018} and having the number of unit cells per supercell \textit{equal} to the number of propagating diffraction orders would demonstrate maximum efficiency of only $70$\% in the shown examples. 

To conclude, in this paper we demonstrate that to perfectly control the diffraction pattern each propagating diffraction order requires two degrees of freedom represented \textit{only} by passive and lossless loaded thin wires. Thus, a metagrating having the number of unit cells per supercell twice the number of propagating diffraction orders allows one to set arbitrary complex amplitudes of all diffracted propagating plane-waves and accurately adjust the near-field in order to satisfy the conditions of passivity and absence of loss. 

Although the proof of concept is done at microwave frequencies under the assumption of TE polarization, the main theoretical result is general.
Significantly decreasing the number of unit cells per wavelength (comparing to metasurfaces) greatly relaxes  the fabrication constraints  what makes it easier to develop metagratings operating at the optical  domain and capable of controlling all propagating diffraction orders. Recently, a metagrating performing perfect  refraction in the first order at mid-infrared frequency range has been fabricated and experimentally tested~\cite{fan2018perfect}.
Control over the reflection  at infrared frequencies was demonstrated in Ref.~\cite{2018arXiv181210164P} by means of numerical simulations.
Meanwhile, presented formulas  can be adapted  for the case of TM polarization (and magnetic line currents) by means of duality relations~\cite{felsen1994radiation,2018arXiv181210164P}.
For example, a unit cell possessing magnetic response can be designed on the basis of a split ring resonator~\cite{Alu2017_metagrating,2018arXiv181210164P}.
Moreover, recent advances in the area of manipulating  acoustic wavefronts~\cite{PhysRevApplied.2.064002,Diaz2018_acoustic,PhysRevB.98.060101,PhysRevApplied.11.014023} suggest that the developed theory can be also generalized for the needs of the acoustics community.

The possibility to develop metagratings operating at different frequency ranges as well as for other domains of physics such as acoustics opens an avenue for a plethora of applications.
Particularly, metagratings can enrich the potential implementations of efficient flat optics 
components and tunable microwave antennas by achieving the benefits of simple excitation, ease of fabrication and integration. 

\section*{acknowledgements}

The authors acknowledge  help of Anil Cheraly  (ONERA) in conducting the experiment.

\appendix

\section{Theory}
\label{app:a}

A single electric line current $\textbf J(\textbf r)=I\delta(y,z)\textbf x_0$  radiates a cylindrical wave with the electric field in the form of Hankel function of the second time zeroth order  $H_0^{(2)}[k\sqrt{y^2+z^2}]$ (see Ref.~\cite{felsen1994radiation})
\begin{equation}\label{eq:ELC_x}
E_x(y,z)=-\frac{k\eta}{4}I H_0^{(2)}[k\sqrt{y^2+z^2}],\quad E_y=E_z=0,
\end{equation}
where $k=\omega\sqrt{\varepsilon\mu}$ and $\eta=\sqrt{\mu/\varepsilon}$.
The electric field created by an infinite array of $N$ equidistant line currents per period $L$ is given by the following series
\begin{eqnarray}
\label{eq:ELCA_x0}
&&E_x(y,z)=-\frac{k\eta}{4}\sum_{q=1}^{N}\sum_{n=-\infty}^\infty I_q e^{-jk\sin[\theta]nL}\nonumber\\
&&\times H_0^{(2)}[k\sqrt{(y-n L-(q-1)d)^2+z^2}],
\end{eqnarray}
the phase $\exp[-jk\sin[\theta]nL]$ appears because of the  plane-wave illumination at angle $\theta$. 
The Poisson's formula applied to the series of Hankel functions ($f(nL)=\exp[-jk\sin[\theta]nL]H_0^{(2)}[k\sqrt{(y-n L-(q-1)d)^2+z^2}]$)
\begin{equation}\label{eq:Poisson_formula}
\sum_{n=-\infty}^{+\infty}f(nL)=\sum_{m=-\infty}^{+\infty}\int_{-\infty}^{+\infty}\frac{dw}{L}f(w)e^{-j\frac{2\pi m}{L}w}.
\end{equation}
is used to express the series~\eqref{eq:ELCA_x0} via plane-waves
\begin{equation}\label{eq:ELCA_x}
E_x(y,z)=-\frac{k\eta}{2L}\sum_{q=1}^N\sum_{m=-\infty}^{\infty}\frac{I_qe^{j\xi_m(q-1)d}}{\B_m}e^{-j\xi_m y-j\B_m |z|}.
\end{equation}
The Fourier transformation of Hanke function is given by the following formula
\begin{equation}\label{eq:FT_H02}
\int\limits_{-\infty}^{+\infty}dw H_0^{(2)}[k\sqrt{(y-w)^2+z^2}]e^{-j\xi_m w}=2\frac{e^{-j\xi_m y-j\B_m|z|}}{\B_m}.
\end{equation}
The magnetic fields corresponding to Eqs.~\eqref{eq:ELC_x} and \eqref{eq:ELCA_x} can be found by means of the Maxwell equations.
The effect of the grounded substrate on the field radiated by the array can be derived  in the same manner as in Ref.~\cite{Epstein2018}. After some algebra one would arrive at equation~\eqref{eq:Am} for the complex amplitudes of propagating and nonpropagating diffraction orders outside the substrate.
The factor $R_m^{TE}$ appearing in the amplitudes corresponds to the Fresnel's reflection coefficient given by the following formula
\begin{eqnarray}
R_m^{TE}=\frac{j\gamma_m^{TE}\tan[\B_m^sh]-1}{j\gamma_m^{TE}\tan[\B_m^sh]+1},\quad \gamma_m^{TE}=\frac{k_s\eta_s\B_m}{k\eta \B_m^s},
\end{eqnarray}
where $\B_s=\sqrt{\E_s \mu_s k^2-\xi_m^2}$ $\eta_s=\eta\sqrt{\mu_s/\E_s}$.

Mutual impedance densities $Z_{qp}^{(m)}$ take into consideration the interaction of the $q^\textup{th}$ wire (located in the zeroth period) with the substrate and adjacent wires and being expressed via the following formulas
\begin{eqnarray}
&&Z_{qp}^{(m)}=\frac{k\eta}{4}\sum_{n=-\infty}^{+\infty} H_0^{(2)}[k|(q-p)d-n L|] e^{-jk\sin[\theta]nL}\nonumber\\
&&+\frac{k\eta}{2L}\sum_{m=-\infty}^{+\infty} e^{j\xi_m (p-q)d}\frac{R_m^{TE}}{\B_m},\quad q\neq p,\nonumber\\
&&Z_{qq}^{(m)}=\frac{k\eta}{2}\sum_{n=1}^{+\infty} \cos[k\sin[\theta]nL]H_0^{(2)}[knL]\nonumber\\
&&+\frac{k\eta}{2L}\sum_{m=-\infty}^{+\infty}\frac{R_m^{TE}}{\B_m}.
\end{eqnarray}
The series containing $R_m^{TE}$ correspond to the interaction with the substrate.
The electric field at the location of the $q^\textup{th}$ wire in the zeroth period created by the rest of $q^\textup{th}$ wires and all other wires ($q\neq p$) is associated with the first terms constituting $Z_{qq}^{(m)}$ and $Z_{qp}^{(m)}$, respectively.

\section{Design procedure and parameters of the experimental samples}\label{app:b}

\begin{figure}[tb]
\includegraphics[width=0.99\linewidth]{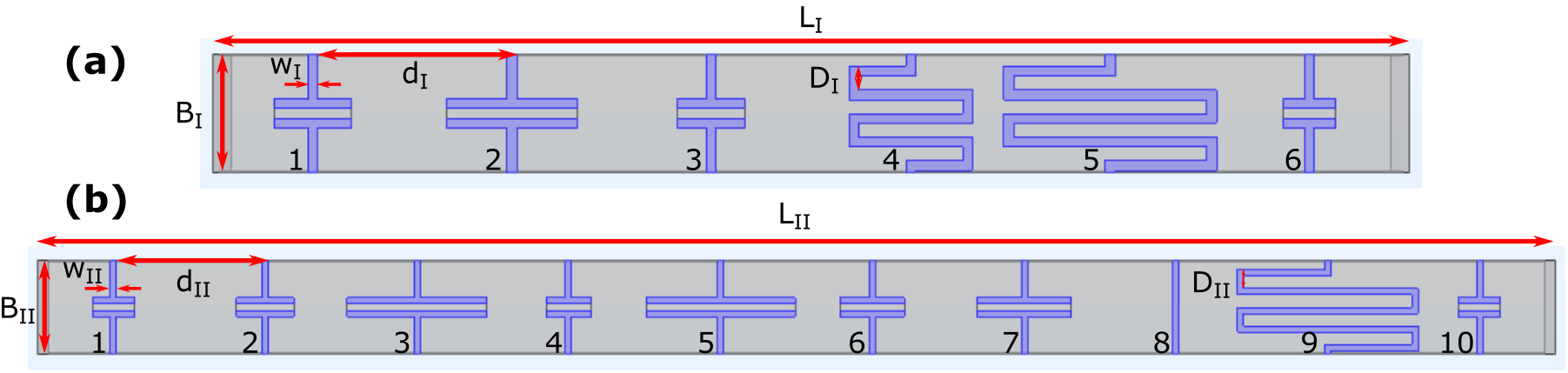}
\caption{\label{fig:s5} 
Outline of the metagratings supercells geometry performing (a) nonspecular reflection at $80$ degrees and (b) beam splitting in the first ($1/3$ of power) and second  ($2/3$\% of power) diffraction orders. Each unit cell of the metagratings is numbered in correspondence with Tab.~\ref{tab:1}.}
\end{figure}

The two metagratings presented as examples in the main text were designed to operate at $10$ GHz ($\lambda\approx 30$ mm).
In order to get the load-impedance densities,  we start by setting the amplitudes of propagating diffraction orders. In the first case of nonspecular reflection of normally incident plane-wave at $80$ degrees, period of the structure is $L_I=30/\sin(80^\textup o)$ mm and there are three propagating diffraction orders $A_{-1}=0$, $A_0=0$ and $A_1=1/\sqrt{\cos(80^\textup{o})}$. It requires six polarization line currents per period  separated by the distance $d_I=L_I/6$. The complex amplitudes of three nonpropagating diffraction orders $A_{-3}$, $A_{-2}$ and $A_{2}$ are found by numerically solving the system of equations (3). After all six amplitudes are known, we calculate the six polarization currents $I_q$ form equation~(1). Then, the load-impedance densities are found from equation~(2). The same procedure is repeated for the other metagrating performing the splitting of normally incident plane-wave between the first ($1/3$ of power) and second ($2/3$ of power) propagating diffraction orders. Period of the metagrating is $L_{II}=2\times 30/\sin(80^\textup o)$ mm and there are ten polarization line currents separated by the distance $d_{II}=L_{II}/10$. The complex amplitudes of the five propagating diffraction orders are set as $A_{-2}=0$, $A_{-1}=0$, $A_0=0$, $A_1=\sqrt{\frac13 /\sqrt{1-(\lambda/L_{II})^2}}$ and $A_2=\sqrt{\frac23/\cos(80^\textup o)}$. Again, the complex  amplitudes of nonpropagating diffraction orders $A_{-5}$, $A_{-4}$, $A_{-3}$, $A_{3}$ and $A_{4}$ are solutions of equation~(3). Computed load-impedance densities can be found in  Table~\ref{tab:1}.

To design experimental samples parameters $w$, $B$ and $D$ are fixed and kept the same for all unit cells in a metagrating, as shown in  Fig.~\ref{fig:s5}. For the first sample performing nonspecular anomalous reflection, $w_I=0.25$ mm, $B_I=3$ mm and $D_I=0.6$ mm. In the case of the second sample used for the beam splitting, these parameters are as follows: $w_{II}=0.25$ mm, $B_{II}=3.75$ mm and $D_{II}=0.75$ mm. The used substrate  is the F4BM220 with $\E_s=2.2(1-j10^{-3})$, $\mu_s=1$, thickness of the substrate is $h=5$ mm. 

In order to find parameters $A$ and $C$ of each unit cell we use equations (4) and (5) presented in the main text and 3D full-wave simulations of a metagrating single supercell (as the ones in   Fig.~\ref{fig:s5}) with imposed periodic boundary conditions. We perform a parametric sweep  with respect to the scaling parameters $\kappa_c$ and $\kappa_i$ until the model acts as desired. For the first and second samples the optimal parameters are $\kappa_c=0.9$, $\kappa_i=1.35$ and  $\kappa_c=0.92$, $\kappa_i=2.66$, respectively. It is important to note that the scaling parameters are independent of the unit cell. In contrast to the design procedure of metasurfaces, here we perform simulations of a whole supercell having $\kappa_c$ and $\kappa_i$ as the only two free parameters. In this way we account for for interaction between different unit cells and immediately arrive at the ultimate design. 
Geometrical parameters of the fabricated samples are specified in  Table~\ref{tab:1}.

\section{Processing of 3D simulation data}\label{app:c}

\begin{figure}[tb]
\includegraphics[width=0.99\linewidth]{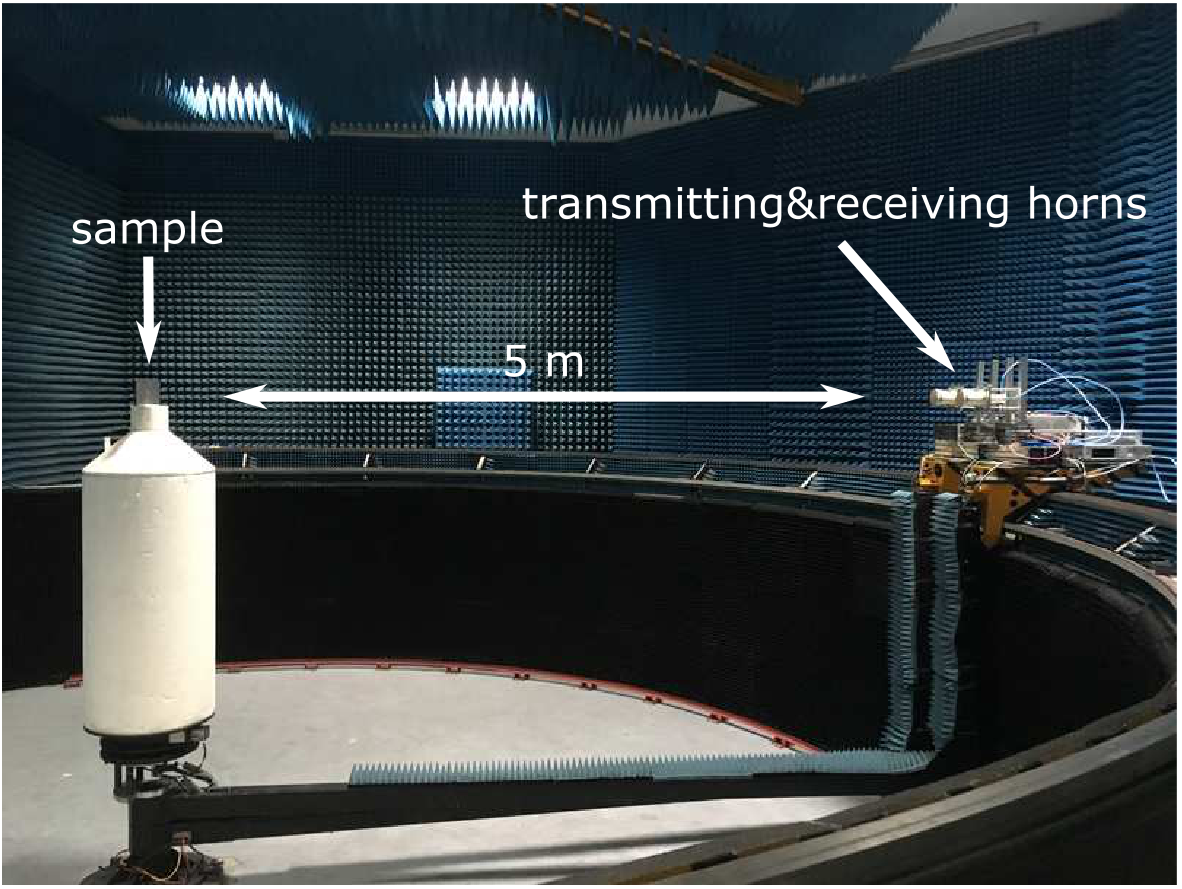}
\caption{\label{fig:s6} 
Photography of the experimental setup used to measure the scattering patterns.}
\end{figure}

\begin{figure}[tb]
\includegraphics[width=0.99\linewidth]{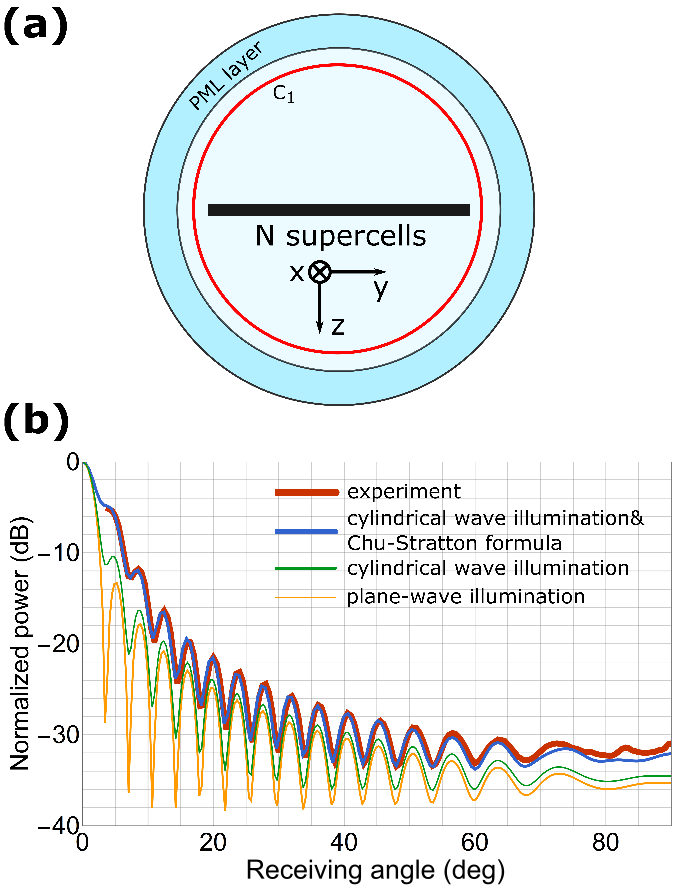}
\caption{\label{fig:s7} 
(a) 2D cross section of the 3D full-wave simulation model of finite size metagrating. The red curve depicts the circle where the scattered fields were extracted. (b) Power scattering pattern from a metallic plate of length $485$ mm simulated numerically under different conditions and compared to the experimental curve, frequency is $10$ GHz.}
\end{figure}

In the measurement setup (see,  Fig.~\ref{fig:s6}), the distance between the antennas and the sample is $5$ m. This distance is not large enough to assume that the measurements are performed under the far-field condition.
Indeed, the physical dimensions of the experimental samples are approximately $480$ mm in the $y$-direction by $160$ mm in the $x$-direction, see photographies in Figs.~\ref{fig:3} (a) and (b). 
Thus the wavefront of the incident wave in the $y$-direction cannot be approximated by a plane-wave. To take it into account, simulations of the finite number of supercells (shown in  Fig.~\ref{fig:s5}) were performed assuming a cylindrical incident wave (phase center is $5$ m away) with periodic boundary conditions applied in the $x$-direction.  In order to correctly  compare  the simulation and measurement results, we harness the Chu-Stratton integration formula~\cite{Chu_Stratton1939,stratton2007electromagnetic} to extrapolate the field calculated on the circle $C_1$ of radius $258.7$ mm (illustrated by the red curve in  Fig.~\ref{fig:s7} (a)) enclosing the sample to the circle $C_2$ with $5$ m radius
\begin{eqnarray}\label{eq:Chu_Stratton}
&&\textbf{E}(y_2,z_2)=\frac{1}{4\pi}\oint_{C_1}\left(i\omega \mu [\textbf{m}\times\textbf{H}(y_1,z_1)]
+[\textbf{m}\times\textbf{E}(y_1,z_1)]\right.\nonumber\\&&\times\nabla 
\left.+[\textbf{m}\textbf{E}(y_1,z_1)]\nabla\right)G(y_2-y_1,z_2-z_1)dl.
\end{eqnarray}
Here $y_2$ and $z_2$ are the coordinates of a point belonging to $C_2$, the integrand contains the fields computed on $C_1$, $G$ is the free space green function and $\textbf m$ is the unit normal vector directed outward $C_1$.  As the simulations are performed with periodic boundary conditions in the $x$-direction, a 2D symmetry is assumed and, thus, we used $G(y,z)=j H_0^{(2)}[k\sqrt{y^2+z^2}]/4$ as a Green function.

Fig.~\ref{fig:s7} (b) demonstrates the importance of  the Chu-Stratton formula. It  compares the scattering patterns from a metallic plate measured experimentally and obtained via numerical simulations under different conditions: (i) the metallic plate is under the normally incident plane-wave, far-field is calculated; (ii) the metallic plate is under the cylindrical wave illumination, phase center is at the distance $5$ m, far-field is calculated; (ii) the metallic plate is under the cylindrical wave illumination, scattered field is processed by means of Chu-Stratton formula and pattern at the distance $5$ m is built.

\section{Calculation of the power scattered in given diffraction order}\label{app:d}

The diffraction pattern appeared when a plane-wave reflects from an infinite metagrating is represented by a finite number of plane-waves propagating at certain angles. 
The power scattered in the $m^\textup{th}$ propagating diffraction order is then calculated as $|A_m^{TE}|^2\beta_m/\beta_0$ (assuming unit amplitude of the incident wave). 
However, when it comes to a finite size periodic structure under a plane-wave like illumination the pattern of the scattered field is much more complex. 
In this case we use the following formula to estimate the part of total power $\alpha_m(\nu)$ scattered in a given diffraction order 
\begin{equation}
\alpha_m(\nu)=\frac{\int_{\theta_{1}^m}^{\theta_{2}^m} P(\nu,\theta)d\theta}{\sum_{m=-l}^r\int_{\theta_{1}^m}^{\theta_{2}^m} P(\nu,\theta)d\theta}.
\end{equation}
Here $P(\nu,\theta)$ represents the power scattered in the receiving angle $\theta$, $\nu$ is the frequency. 
The integration is performed only over the receiving angle range of half the maximum power of the beam corresponding to the  $m^\textup{th}$ diffraction order. 
The summation in the denominator includes all propagating diffraction orders at the frequency $\nu$. 
Angles $\theta_1^m$ and $\theta_2^m$ are found as follows. 
First, we accurately localize the maximum of the $m^\textup{th}$ diffraction order around the receiving angle $\sin^{-1}(\xi_m/k)$. 
Then, $\theta_1^m$ and $\theta_2^m$ correspond to the  $-3$ dB of the power attenuation from the found maximum value.

\section{2D full-wave simulations of metagratings with COMSOL Multiphysics}\label{app:e}

\begin{figure}[tb]
\includegraphics[width=0.99\linewidth]{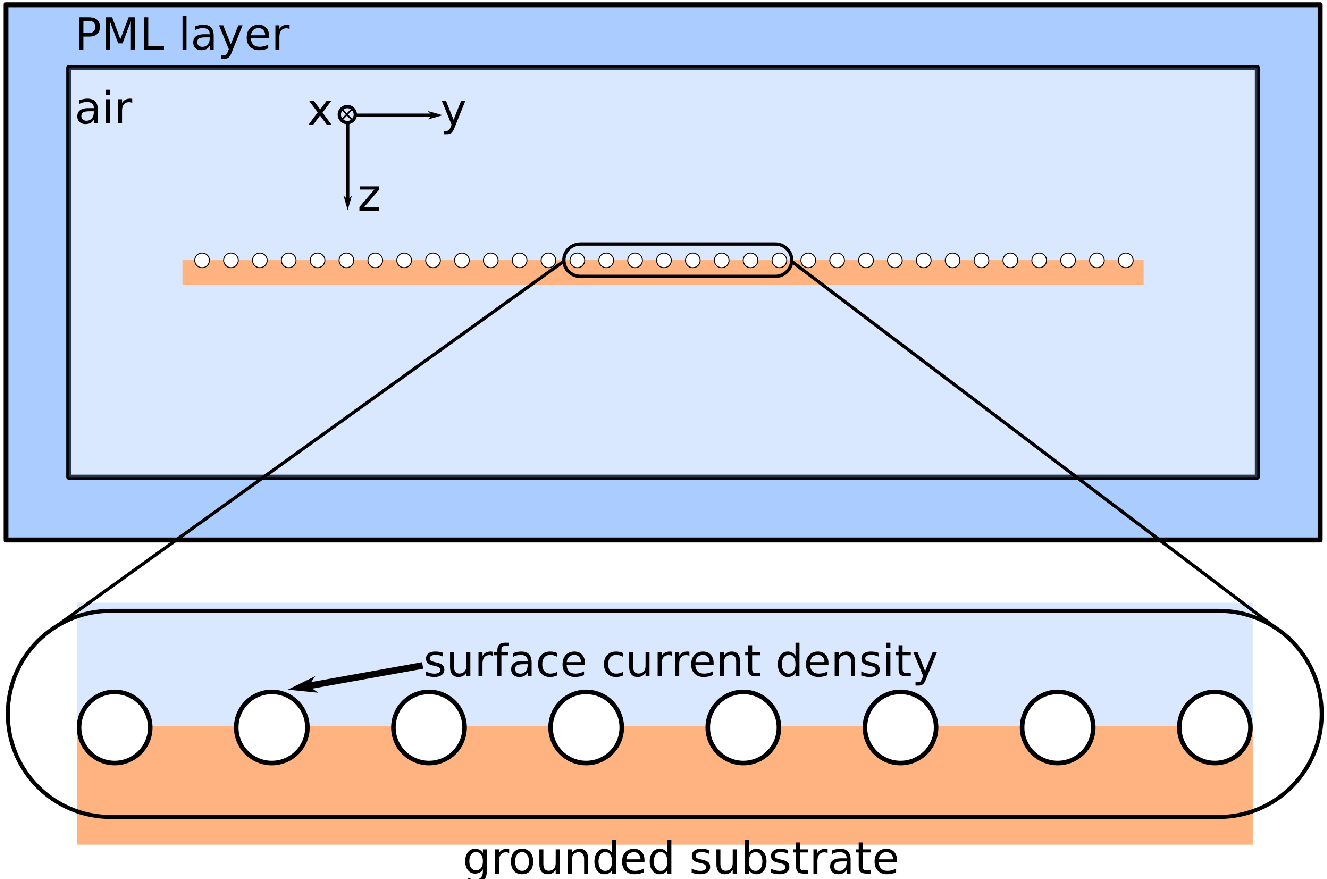}
\caption{\label{fig:s8} 
Schematics of the 2D COMSOL model used for simulating metagratings. The white regions inside the circles are excluded from the model. Polarization line currents of effective radius $r_0$ (radius of the circles, it represents the input-impedance density) are simulated as surface current density: $\textbf J_{es}=E_x/Z_q/(2\pi r_0)\textbf x_0$ ($\textbf x_0$ is the unit vector in the $x$ direction). The total number of line currents (circles) is number of line currents per period times the number of periods.}
\end{figure}

A  COMSOL model for 2D full-wave simulations of metagratings can be built in the following way. The principal element of a metagrating is a polarization line current which is modeled in COMSOL as surface current density assigned to the boundary of a circle, as shown in Fig.~\ref{fig:s8}. The radius of the circle $r_0$ should be equal to the effective radius of a thin wire in order to get the correct value of the input-impedance density. It is important to exclude from the model the interior of the circles, otherwise one would get an incorrect value of the input-impedance density. The surface current density $\textbf J_{es}$ is set as follows: $E_x/Z_q/(2\pi r_0)\textbf x_0$ ($Z_q$ is the load-impedance density of the $q^\textup{th}$ thin wire). The array of circles is placed on a PEC-backed substrate (the circles' centers are on the top of the substrate) as shown in  Fig.~\ref{fig:s8}. In order to excite the model we use scattered field formulation and set a background field. The rest of the model is standard and can be understood from  Fig.~\ref{fig:s8}.


\begin{thebibliography}{47}%
	\makeatletter
	\providecommand \@ifxundefined [1]{%
		\@ifx{#1\undefined}
	}%
	\providecommand \@ifnum [1]{%
		\ifnum #1\expandafter \@firstoftwo
		\else \expandafter \@secondoftwo
		\fi
	}%
	\providecommand \@ifx [1]{%
		\ifx #1\expandafter \@firstoftwo
		\else \expandafter \@secondoftwo
		\fi
	}%
	\providecommand \natexlab [1]{#1}%
	\providecommand \enquote  [1]{``#1''}%
	\providecommand \bibnamefont  [1]{#1}%
	\providecommand \bibfnamefont [1]{#1}%
	\providecommand \citenamefont [1]{#1}%
	\providecommand \href@noop [0]{\@secondoftwo}%
	\providecommand \href [0]{\begingroup \@sanitize@url \@href}%
	\providecommand \@href[1]{\@@startlink{#1}\@@href}%
	\providecommand \@@href[1]{\endgroup#1\@@endlink}%
	\providecommand \@sanitize@url [0]{\catcode `\\12\catcode `\$12\catcode
		`\&12\catcode `\#12\catcode `\^12\catcode `\_12\catcode `\%12\relax}%
	\providecommand \@@startlink[1]{}%
	\providecommand \@@endlink[0]{}%
	\providecommand \url  [0]{\begingroup\@sanitize@url \@url }%
	\providecommand \@url [1]{\endgroup\@href {#1}{\urlprefix }}%
	\providecommand \urlprefix  [0]{URL }%
	\providecommand \Eprint [0]{\href }%
	\providecommand \doibase [0]{http://dx.doi.org/}%
	\providecommand \selectlanguage [0]{\@gobble}%
	\providecommand \bibinfo  [0]{\@secondoftwo}%
	\providecommand \bibfield  [0]{\@secondoftwo}%
	\providecommand \translation [1]{[#1]}%
	\providecommand \BibitemOpen [0]{}%
	\providecommand \bibitemStop [0]{}%
	\providecommand \bibitemNoStop [0]{.\EOS\space}%
	\providecommand \EOS [0]{\spacefactor3000\relax}%
	\providecommand \BibitemShut  [1]{\csname bibitem#1\endcsname}%
	\let\auto@bib@innerbib\@empty
	\bibitem [{\citenamefont {Wood}(1910)}]{Wood1910}%
	\BibitemOpen
	\bibfield  {author} {\bibinfo {author} {\bibfnamefont {R.W.}\ \bibnamefont
			{Wood}},\ }\bibfield  {title} {\enquote {\bibinfo {title} {The echelette
				grating for the infra-red},}\ }\href {\doibase 10.1080/14786441008636964}
	{\bibfield  {journal} {\bibinfo  {journal} {The London, Edinburgh, and Dublin
				Philosophical Magazine and Journal of Science}\ }\textbf {\bibinfo {volume}
			{20}},\ \bibinfo {pages} {770--778} (\bibinfo {year} {1910})},\ \Eprint
	{http://arxiv.org/abs/https://doi.org/10.1080/14786441008636964}
	{https://doi.org/10.1080/14786441008636964} \BibitemShut {NoStop}%
	\bibitem [{\citenamefont {Rowland}(1893)}]{Rowland1893}%
	\BibitemOpen
	\bibfield  {author} {\bibinfo {author} {\bibfnamefont {Henry~A.}\
			\bibnamefont {Rowland}},\ }\bibfield  {title} {\enquote {\bibinfo {title}
			{Gratings in theory and practice},}\ }\href {\doibase
		10.1080/14786449308620425} {\bibfield  {journal} {\bibinfo  {journal} {The
				London, Edinburgh, and Dublin Philosophical Magazine and Journal of Science}\
		}\textbf {\bibinfo {volume} {35}},\ \bibinfo {pages} {397--419} (\bibinfo
		{year} {1893})},\ \Eprint
	{http://arxiv.org/abs/https://doi.org/10.1080/14786449308620425}
	{https://doi.org/10.1080/14786449308620425} \BibitemShut {NoStop}%
	\bibitem [{\citenamefont {Stamm}\ and\ \citenamefont
		{Whalen}(1946)}]{stamm1946energy}%
	\BibitemOpen
	\bibfield  {author} {\bibinfo {author} {\bibfnamefont {R.~F.}\ \bibnamefont
			{Stamm}}\ and\ \bibinfo {author} {\bibfnamefont {J.~J.}\ \bibnamefont
			{Whalen}},\ }\bibfield  {title} {\enquote {\bibinfo {title} {Energy
				distribution of diffraction gratings as a function of groove form
				(calculations by an equation of henry a. rowland)},}\ }\href@noop {}
	{\bibfield  {journal} {\bibinfo  {journal} {JOSA}\ }\textbf {\bibinfo
			{volume} {36}},\ \bibinfo {pages} {2--12} (\bibinfo {year}
		{1946})}\BibitemShut {NoStop}%
	\bibitem [{\citenamefont {Breidne}\ \emph {et~al.}(1979)\citenamefont
		{Breidne}, \citenamefont {Johansson}, \citenamefont {Nilsson},\ and\
		\citenamefont {\r{A}hl\`en}}]{Breidne1979}%
	\BibitemOpen
	\bibfield  {author} {\bibinfo {author} {\bibfnamefont {M.}~\bibnamefont
			{Breidne}}, \bibinfo {author} {\bibfnamefont {S.}~\bibnamefont {Johansson}},
		\bibinfo {author} {\bibfnamefont {L-E.}\ \bibnamefont {Nilsson}}, \ and\
		\bibinfo {author} {\bibfnamefont {H.}~\bibnamefont {\r{A}hl\`en}},\
	}\bibfield  {title} {\enquote {\bibinfo {title} {Blazed holographic
				gratings},}\ }\href {\doibase 10.1080/713819919} {\bibfield  {journal}
		{\bibinfo  {journal} {Optica Acta: International Journal of Optics}\ }\textbf
		{\bibinfo {volume} {26}},\ \bibinfo {pages} {1427--1441} (\bibinfo {year}
		{1979})},\ \Eprint {http://arxiv.org/abs/https://doi.org/10.1080/713819919}
	{https://doi.org/10.1080/713819919} \BibitemShut {NoStop}%
	\bibitem [{\citenamefont {Hessel}\ \emph {et~al.}(1975)\citenamefont {Hessel},
		\citenamefont {Schmoys},\ and\ \citenamefont {Tseng}}]{Hessel1975}%
	\BibitemOpen
	\bibfield  {author} {\bibinfo {author} {\bibfnamefont {A.}~\bibnamefont
			{Hessel}}, \bibinfo {author} {\bibfnamefont {J.}~\bibnamefont {Schmoys}}, \
		and\ \bibinfo {author} {\bibfnamefont {D.~Y.}\ \bibnamefont {Tseng}},\
	}\bibfield  {title} {\enquote {\bibinfo {title} {Bragg-angle blazing of
				diffraction gratings$\ast$},}\ }\href {\doibase 10.1364/JOSA.65.000380}
	{\bibfield  {journal} {\bibinfo  {journal} {J. Opt. Soc. Am.}\ }\textbf
		{\bibinfo {volume} {65}},\ \bibinfo {pages} {380--384} (\bibinfo {year}
		{1975})}\BibitemShut {NoStop}%
	\bibitem [{\citenamefont {Breidne}\ and\ \citenamefont
		{Maystre}(1981)}]{Breidne1981}%
	\BibitemOpen
	\bibfield  {author} {\bibinfo {author} {\bibfnamefont {M.}~\bibnamefont
			{Breidne}}\ and\ \bibinfo {author} {\bibfnamefont {D.}~\bibnamefont
			{Maystre}},\ }\bibfield  {title} {\enquote {\bibinfo {title} {Perfect blaze
				in non-littrow mountings},}\ }\href {\doibase 10.1080/713820450} {\bibfield
		{journal} {\bibinfo  {journal} {Optica Acta: International Journal of
				Optics}\ }\textbf {\bibinfo {volume} {28}},\ \bibinfo {pages} {1321--1327}
		(\bibinfo {year} {1981})},\ \Eprint
	{http://arxiv.org/abs/https://doi.org/10.1080/713820450}
	{https://doi.org/10.1080/713820450} \BibitemShut {NoStop}%
	\bibitem [{\citenamefont {Pozar}(1996)}]{Pozar1996}%
	\BibitemOpen
	\bibfield  {author} {\bibinfo {author} {\bibfnamefont {D.~M.}\ \bibnamefont
			{Pozar}},\ }\bibfield  {title} {\enquote {\bibinfo {title} {Flat lens antenna
				concept using aperture coupled microstrip patches},}\ }\href {\doibase
		10.1049/el:19961451} {\bibfield  {journal} {\bibinfo  {journal} {Electronics
				Letters}\ }\textbf {\bibinfo {volume} {32}},\ \bibinfo {pages} {2109--2111}
		(\bibinfo {year} {1996})}\BibitemShut {NoStop}%
	\bibitem [{\citenamefont {Huang}\ and\ \citenamefont
		{Encinar}(2007)}]{huang2007reflectarray}%
	\BibitemOpen
	\bibfield  {author} {\bibinfo {author} {\bibfnamefont {J.}~\bibnamefont
			{Huang}}\ and\ \bibinfo {author} {\bibfnamefont {J.~A.}\ \bibnamefont
			{Encinar}},\ }\href@noop {} {\emph {\bibinfo {title} {Reflectarray
				antennas}}},\ Vol.~\bibinfo {volume} {30}\ (\bibinfo  {publisher} {John Wiley
		\& Sons, Hoboken, New Jersey},\ \bibinfo {year} {2007})\BibitemShut {NoStop}%
	\bibitem [{\citenamefont {Pozar}(2007)}]{Pozar2007}%
	\BibitemOpen
	\bibfield  {author} {\bibinfo {author} {\bibfnamefont {D.~M.}\ \bibnamefont
			{Pozar}},\ }\bibfield  {title} {\enquote {\bibinfo {title} {Wideband
				reflectarrays using artificial impedance surfaces},}\ }\href@noop {}
	{\bibfield  {journal} {\bibinfo  {journal} {Electronics Letters}\ }\textbf
		{\bibinfo {volume} {43}},\ \bibinfo {pages} {1--2} (\bibinfo {year}
		{2007})}\BibitemShut {NoStop}%
	\bibitem [{\citenamefont {Jahani}\ and\ \citenamefont
		{Jacob}(2016)}]{Jacob2016}%
	\BibitemOpen
	\bibfield  {author} {\bibinfo {author} {\bibfnamefont {S.}~\bibnamefont
			{Jahani}}\ and\ \bibinfo {author} {\bibfnamefont {Z.}~\bibnamefont {Jacob}},\
	}\bibfield  {title} {\enquote {\bibinfo {title} {All-dielectric
				metamaterials},}\ }\href {\doibase 10.1038/nnano.2015.304} {\bibfield
		{journal} {\bibinfo  {journal} {Nat. Nanotech.}\ }\textbf {\bibinfo {volume}
			{11}},\ \bibinfo {pages} {23--36} (\bibinfo {year} {2016})}\BibitemShut
	{NoStop}%
	\bibitem [{\citenamefont {Glybovski}\ \emph {et~al.}(2016)\citenamefont
		{Glybovski}, \citenamefont {Tretyakov}, \citenamefont {Belov}, \citenamefont
		{Kivshar},\ and\ \citenamefont {Simovski}}]{Glybovski2016}%
	\BibitemOpen
	\bibfield  {author} {\bibinfo {author} {\bibfnamefont {S.~B.}\ \bibnamefont
			{Glybovski}}, \bibinfo {author} {\bibfnamefont {S.~A.}\ \bibnamefont
			{Tretyakov}}, \bibinfo {author} {\bibfnamefont {P.~A.}\ \bibnamefont
			{Belov}}, \bibinfo {author} {\bibfnamefont {Y.~S.}\ \bibnamefont {Kivshar}},
		\ and\ \bibinfo {author} {\bibfnamefont {C.~R.}\ \bibnamefont {Simovski}},\
	}\bibfield  {title} {\enquote {\bibinfo {title} {{Metasurfaces: From
					microwaves to visible}},}\ }\href {\doibase 10.1016/j.physrep.2016.04.004}
	{\bibfield  {journal} {\bibinfo  {journal} {Physics Reports}\ }\textbf
		{\bibinfo {volume} {634}},\ \bibinfo {pages} {1--72} (\bibinfo {year}
		{2016})}\BibitemShut {NoStop}%
	\bibitem [{\citenamefont {Sun}\ \emph {et~al.}(2018)\citenamefont {Sun},
		\citenamefont {Fan}, \citenamefont {Zhang}, \citenamefont {Zhang},
		\citenamefont {Shi}, \citenamefont {Wang}, \citenamefont {Xie}, \citenamefont
		{Wang}, \citenamefont {Fan}, \citenamefont {Liu}, \citenamefont {Liu},
		\citenamefont {Li}, \citenamefont {Yan},\ and\ \citenamefont
		{Guo}}]{C7TC03384B}%
	\BibitemOpen
	\bibfield  {author} {\bibinfo {author} {\bibfnamefont {K.}~\bibnamefont
			{Sun}}, \bibinfo {author} {\bibfnamefont {R.}~\bibnamefont {Fan}}, \bibinfo
		{author} {\bibfnamefont {X.}~\bibnamefont {Zhang}}, \bibinfo {author}
		{\bibfnamefont {Z.}~\bibnamefont {Zhang}}, \bibinfo {author} {\bibfnamefont
			{Z.}~\bibnamefont {Shi}}, \bibinfo {author} {\bibfnamefont {N.}~\bibnamefont
			{Wang}}, \bibinfo {author} {\bibfnamefont {P.}~\bibnamefont {Xie}}, \bibinfo
		{author} {\bibfnamefont {Z.}~\bibnamefont {Wang}}, \bibinfo {author}
		{\bibfnamefont {G.}~\bibnamefont {Fan}}, \bibinfo {author} {\bibfnamefont
			{H.}~\bibnamefont {Liu}}, \bibinfo {author} {\bibfnamefont {C.}~\bibnamefont
			{Liu}}, \bibinfo {author} {\bibfnamefont {T.}~\bibnamefont {Li}}, \bibinfo
		{author} {\bibfnamefont {C.}~\bibnamefont {Yan}}, \ and\ \bibinfo {author}
		{\bibfnamefont {Z.}~\bibnamefont {Guo}},\ }\bibfield  {title} {\enquote
		{\bibinfo {title} {An overview of metamaterials and their achievements in
				wireless power transfer},}\ }\href {\doibase 10.1039/C7TC03384B} {\bibfield
		{journal} {\bibinfo  {journal} {J. Mater. Chem. C}\ }\textbf {\bibinfo
			{volume} {6}},\ \bibinfo {pages} {2925--2943} (\bibinfo {year}
		{2018})}\BibitemShut {NoStop}%
	\bibitem [{\citenamefont {Tong}(2018)}]{Tong2018}%
	\BibitemOpen
	\bibfield  {author} {\bibinfo {author} {\bibfnamefont {X.C.}\ \bibnamefont
			{Tong}},\ }\href@noop {} {\emph {\bibinfo {title} {Functional Metamaterials
				and Metadevices}}}\ (\bibinfo  {publisher} {Cham, Springer},\ \bibinfo {year}
	{2018})\BibitemShut {NoStop}%
	\bibitem [{\citenamefont {Yu}\ \emph {et~al.}(2011)\citenamefont {Yu},
		\citenamefont {Genevet}, \citenamefont {Kats}, \citenamefont {Aieta},
		\citenamefont {Tetienne}, \citenamefont {Capasso},\ and\ \citenamefont
		{Gaburro}}]{Capasso_GeneralizedReflectionLaw}%
	\BibitemOpen
	\bibfield  {author} {\bibinfo {author} {\bibfnamefont {N.}~\bibnamefont
			{Yu}}, \bibinfo {author} {\bibfnamefont {P.}~\bibnamefont {Genevet}},
		\bibinfo {author} {\bibfnamefont {M.~A.}\ \bibnamefont {Kats}}, \bibinfo
		{author} {\bibfnamefont {F.}~\bibnamefont {Aieta}}, \bibinfo {author}
		{\bibfnamefont {J.-P.}\ \bibnamefont {Tetienne}}, \bibinfo {author}
		{\bibfnamefont {F.}~\bibnamefont {Capasso}}, \ and\ \bibinfo {author}
		{\bibfnamefont {Z.}~\bibnamefont {Gaburro}},\ }\bibfield  {title} {\enquote
		{\bibinfo {title} {Light propagation with phase discontinuities: Generalized
				laws of reflection and refraction},}\ }\href {\doibase
		10.1126/science.1210713} {\bibfield  {journal} {\bibinfo  {journal}
			{Science}\ }\textbf {\bibinfo {volume} {334}},\ \bibinfo {pages} {333--337}
		(\bibinfo {year} {2011})}\BibitemShut {NoStop}%
	\bibitem [{\citenamefont {Pfeiffer}\ and\ \citenamefont
		{Grbic}(2013)}]{Pfeiffer2013}%
	\BibitemOpen
	\bibfield  {author} {\bibinfo {author} {\bibfnamefont {C.}~\bibnamefont
			{Pfeiffer}}\ and\ \bibinfo {author} {\bibfnamefont {A.}~\bibnamefont
			{Grbic}},\ }\bibfield  {title} {\enquote {\bibinfo {title} {{Metamaterial
					Huygens' surfaces: Tailoring wave fronts with reflectionless sheets}},}\
	}\href {\doibase 10.1103/PhysRevLett.110.197401} {\bibfield  {journal}
		{\bibinfo  {journal} {Physical Review Letters}\ }\textbf {\bibinfo {volume}
			{110}},\ \bibinfo {pages} {1--5} (\bibinfo {year} {2013})}\BibitemShut
	{NoStop}%
	\bibitem [{\citenamefont {Asadchy}\ \emph {et~al.}(2016)\citenamefont
		{Asadchy}, \citenamefont {Albooyeh}, \citenamefont {Tcvetkova}, \citenamefont
		{D\'{\i}az-Rubio}, \citenamefont {Ra'di},\ and\ \citenamefont
		{Tretyakov}}]{Asadchy2016_SpatiallyDispMS}%
	\BibitemOpen
	\bibfield  {author} {\bibinfo {author} {\bibfnamefont {V.~S.}\ \bibnamefont
			{Asadchy}}, \bibinfo {author} {\bibfnamefont {M.}~\bibnamefont {Albooyeh}},
		\bibinfo {author} {\bibfnamefont {S.~N.}\ \bibnamefont {Tcvetkova}}, \bibinfo
		{author} {\bibfnamefont {A.}~\bibnamefont {D\'{\i}az-Rubio}}, \bibinfo
		{author} {\bibfnamefont {Y.}~\bibnamefont {Ra'di}}, \ and\ \bibinfo {author}
		{\bibfnamefont {S.~A.}\ \bibnamefont {Tretyakov}},\ }\bibfield  {title}
	{\enquote {\bibinfo {title} {Perfect control of reflection and refraction
				using spatially dispersive metasurfaces},}\ }\href {\doibase
		10.1103/PhysRevB.94.075142} {\bibfield  {journal} {\bibinfo  {journal} {Phys.
				Rev. B}\ }\textbf {\bibinfo {volume} {94}},\ \bibinfo {pages} {075142}
		(\bibinfo {year} {2016})}\BibitemShut {NoStop}%
	\bibitem [{\citenamefont {Asadchy}\ \emph {et~al.}(2017)\citenamefont
		{Asadchy}, \citenamefont {D\'{\i}az-Rubio}, \citenamefont {Tcvetkova},
		\citenamefont {Kwon}, \citenamefont {Elsakka}, \citenamefont {Albooyeh},\
		and\ \citenamefont {Tretyakov}}]{Asadchy2017MultiChannel}%
	\BibitemOpen
	\bibfield  {author} {\bibinfo {author} {\bibfnamefont {V.~S.}\ \bibnamefont
			{Asadchy}}, \bibinfo {author} {\bibfnamefont {A.}~\bibnamefont
			{D\'{\i}az-Rubio}}, \bibinfo {author} {\bibfnamefont {S.~N.}\ \bibnamefont
			{Tcvetkova}}, \bibinfo {author} {\bibfnamefont {D.-H.}\ \bibnamefont {Kwon}},
		\bibinfo {author} {\bibfnamefont {A.}~\bibnamefont {Elsakka}}, \bibinfo
		{author} {\bibfnamefont {M.}~\bibnamefont {Albooyeh}}, \ and\ \bibinfo
		{author} {\bibfnamefont {S.~A.}\ \bibnamefont {Tretyakov}},\ }\bibfield
	{title} {\enquote {\bibinfo {title} {Flat engineered multichannel
				reflectors},}\ }\href {\doibase 10.1103/PhysRevX.7.031046} {\bibfield
		{journal} {\bibinfo  {journal} {Phys. Rev. X}\ }\textbf {\bibinfo {volume}
			{7}},\ \bibinfo {pages} {031046} (\bibinfo {year} {2017})}\BibitemShut
	{NoStop}%
	\bibitem [{\citenamefont {Epstein}\ and\ \citenamefont
		{Eleftheriades}(2016{\natexlab{a}})}]{Epstein2016_fieldTrans_OBMS}%
	\BibitemOpen
	\bibfield  {author} {\bibinfo {author} {\bibfnamefont {A.}~\bibnamefont
			{Epstein}}\ and\ \bibinfo {author} {\bibfnamefont {G.~V.}\ \bibnamefont
			{Eleftheriades}},\ }\bibfield  {title} {\enquote {\bibinfo {title} {Arbitrary
				power-conserving field transformations with passive lossless omega-type
				bianisotropic metasurfaces},}\ }\href {\doibase 10.1109/TAP.2016.2588495}
	{\bibfield  {journal} {\bibinfo  {journal} {IEEE Transactions on Antennas and
				Propagation}\ }\textbf {\bibinfo {volume} {64}},\ \bibinfo {pages}
		{3880--3895} (\bibinfo {year} {2016}{\natexlab{a}})}\BibitemShut {NoStop}%
	\bibitem [{\citenamefont {Chen}\ \emph {et~al.}(2018)\citenamefont {Chen},
		\citenamefont {Abdo-S\'anchez}, \citenamefont {Epstein},\ and\ \citenamefont
		{Eleftheriades}}]{Epstein2018_exp_anrefr}%
	\BibitemOpen
	\bibfield  {author} {\bibinfo {author} {\bibfnamefont {M.}~\bibnamefont
			{Chen}}, \bibinfo {author} {\bibfnamefont {E.}~\bibnamefont
			{Abdo-S\'anchez}}, \bibinfo {author} {\bibfnamefont {A.}~\bibnamefont
			{Epstein}}, \ and\ \bibinfo {author} {\bibfnamefont {G.~V.}\ \bibnamefont
			{Eleftheriades}},\ }\bibfield  {title} {\enquote {\bibinfo {title} {Theory,
				design, and experimental verification of a reflectionless bianisotropic
				huygens' metasurface for wide-angle refraction},}\ }\href {\doibase
		10.1103/PhysRevB.97.125433} {\bibfield  {journal} {\bibinfo  {journal} {Phys.
				Rev. B}\ }\textbf {\bibinfo {volume} {97}},\ \bibinfo {pages} {125433}
		(\bibinfo {year} {2018})}\BibitemShut {NoStop}%
	\bibitem [{\citenamefont {Lavigne}\ \emph {et~al.}(2018)\citenamefont
		{Lavigne}, \citenamefont {Achouri}, \citenamefont {Asadchy}, \citenamefont
		{Tretyakov},\ and\ \citenamefont {Caloz}}]{8259235}%
	\BibitemOpen
	\bibfield  {author} {\bibinfo {author} {\bibfnamefont {G.}~\bibnamefont
			{Lavigne}}, \bibinfo {author} {\bibfnamefont {K.}~\bibnamefont {Achouri}},
		\bibinfo {author} {\bibfnamefont {V.~S.}\ \bibnamefont {Asadchy}}, \bibinfo
		{author} {\bibfnamefont {S.~A.}\ \bibnamefont {Tretyakov}}, \ and\ \bibinfo
		{author} {\bibfnamefont {C.}~\bibnamefont {Caloz}},\ }\bibfield  {title}
	{\enquote {\bibinfo {title} {Susceptibility derivation and experimental
				demonstration of refracting metasurfaces without spurious diffraction},}\
	}\href {\doibase 10.1109/TAP.2018.2793958} {\bibfield  {journal} {\bibinfo
			{journal} {IEEE Transactions on Antennas and Propagation}\ }\textbf {\bibinfo
			{volume} {66}},\ \bibinfo {pages} {1321--1330} (\bibinfo {year}
		{2018})}\BibitemShut {NoStop}%
	\bibitem [{\citenamefont {Epstein}\ and\ \citenamefont
		{Eleftheriades}(2016{\natexlab{b}})}]{Epstein2016_AuxiliryFields}%
	\BibitemOpen
	\bibfield  {author} {\bibinfo {author} {\bibfnamefont {A.}~\bibnamefont
			{Epstein}}\ and\ \bibinfo {author} {\bibfnamefont {G.~V.}\ \bibnamefont
			{Eleftheriades}},\ }\bibfield  {title} {\enquote {\bibinfo {title} {Synthesis
				of passive lossless metasurfaces using auxiliary fields for reflectionless
				beam splitting and perfect reflection},}\ }\href {\doibase
		10.1103/PhysRevLett.117.256103} {\bibfield  {journal} {\bibinfo  {journal}
			{Phys. Rev. Lett.}\ }\textbf {\bibinfo {volume} {117}},\ \bibinfo {pages}
		{256103} (\bibinfo {year} {2016}{\natexlab{b}})}\BibitemShut {NoStop}%
	\bibitem [{\citenamefont {Kwon}\ and\ \citenamefont
		{Tretyakov}(2017)}]{Tretyakov_2017_auxilirySW}%
	\BibitemOpen
	\bibfield  {author} {\bibinfo {author} {\bibfnamefont {D.-H.}\ \bibnamefont
			{Kwon}}\ and\ \bibinfo {author} {\bibfnamefont {S.~A.}\ \bibnamefont
			{Tretyakov}},\ }\bibfield  {title} {\enquote {\bibinfo {title} {Perfect
				reflection control for impenetrable surfaces using surface waves of
				orthogonal polarization},}\ }\href {\doibase 10.1103/PhysRevB.96.085438}
	{\bibfield  {journal} {\bibinfo  {journal} {Phys. Rev. B}\ }\textbf {\bibinfo
			{volume} {96}},\ \bibinfo {pages} {085438} (\bibinfo {year}
		{2017})}\BibitemShut {NoStop}%
	\bibitem [{\citenamefont {D{\'\i}az-Rubio}\ \emph {et~al.}(2017)\citenamefont
		{D{\'\i}az-Rubio}, \citenamefont {Asadchy}, \citenamefont {Elsakka},\ and\
		\citenamefont {Tretyakov}}]{Tretyakov2017_perfectAR}%
	\BibitemOpen
	\bibfield  {author} {\bibinfo {author} {\bibfnamefont {A.}~\bibnamefont
			{D{\'\i}az-Rubio}}, \bibinfo {author} {\bibfnamefont {V.~S.}\ \bibnamefont
			{Asadchy}}, \bibinfo {author} {\bibfnamefont {A.}~\bibnamefont {Elsakka}}, \
		and\ \bibinfo {author} {\bibfnamefont {S.~A.}\ \bibnamefont {Tretyakov}},\
	}\bibfield  {title} {\enquote {\bibinfo {title} {From the generalized
				reflection law to the realization of perfect anomalous reflectors},}\ }\href
	{\doibase 10.1126/sciadv.1602714} {\bibfield  {journal} {\bibinfo  {journal}
			{Science Advances}\ }\textbf {\bibinfo {volume} {3}} (\bibinfo {year}
		{2017}),\ 10.1126/sciadv.1602714}\BibitemShut {NoStop}%
	\bibitem [{\citenamefont {Kwon}(2018)}]{Kwon2018}%
	\BibitemOpen
	\bibfield  {author} {\bibinfo {author} {\bibfnamefont {D.}~\bibnamefont
			{Kwon}},\ }\bibfield  {title} {\enquote {\bibinfo {title} {Lossless scalar
				metasurfaces for anomalous reflection based on efficient surface field
				optimization},}\ }\href {\doibase 10.1109/LAWP.2018.2836299} {\bibfield
		{journal} {\bibinfo  {journal} {IEEE Antennas and Wireless Propagation
				Letters}\ }\textbf {\bibinfo {volume} {17}},\ \bibinfo {pages} {1149--1152}
		(\bibinfo {year} {2018})}\BibitemShut {NoStop}%
	\bibitem [{\citenamefont {Monticone}\ \emph {et~al.}(2013)\citenamefont
		{Monticone}, \citenamefont {Estakhri},\ and\ \citenamefont
		{Al\`u}}]{PhysRevLett.110.203903}%
	\BibitemOpen
	\bibfield  {author} {\bibinfo {author} {\bibfnamefont {F.}~\bibnamefont
			{Monticone}}, \bibinfo {author} {\bibfnamefont {N.~M.}\ \bibnamefont
			{Estakhri}}, \ and\ \bibinfo {author} {\bibfnamefont {A.}~\bibnamefont
			{Al\`u}},\ }\bibfield  {title} {\enquote {\bibinfo {title} {Full control of
				nanoscale optical transmission with a composite metascreen},}\ }\href
	{\doibase 10.1103/PhysRevLett.110.203903} {\bibfield  {journal} {\bibinfo
			{journal} {Phys. Rev. Lett.}\ }\textbf {\bibinfo {volume} {110}},\ \bibinfo
		{pages} {203903} (\bibinfo {year} {2013})}\BibitemShut {NoStop}%
	\bibitem [{\citenamefont {Epstein}\ and\ \citenamefont
		{Eleftheriades}(2014)}]{Epstein2014_HMS_diffraction}%
	\BibitemOpen
	\bibfield  {author} {\bibinfo {author} {\bibfnamefont {A.}~\bibnamefont
			{Epstein}}\ and\ \bibinfo {author} {\bibfnamefont {G.~V.}\ \bibnamefont
			{Eleftheriades}},\ }\bibfield  {title} {\enquote {\bibinfo {title}
			{{Floquet-Bloch analysis of refracting Huygens metasurfaces}},}\ }\href
	{\doibase 10.1103/PhysRevB.90.235127} {\bibfield  {journal} {\bibinfo
			{journal} {Physical Review B - Condensed Matter and Materials Physics}\
		}\textbf {\bibinfo {volume} {90}},\ \bibinfo {pages} {1--10} (\bibinfo {year}
		{2014})}\BibitemShut {NoStop}%
	\bibitem [{\citenamefont {Ra'di}\ \emph {et~al.}(2017)\citenamefont {Ra'di},
		\citenamefont {Sounas},\ and\ \citenamefont {Al\`u}}]{Alu2017_metagrating}%
	\BibitemOpen
	\bibfield  {author} {\bibinfo {author} {\bibfnamefont {Y.}~\bibnamefont
			{Ra'di}}, \bibinfo {author} {\bibfnamefont {D.~L.}\ \bibnamefont {Sounas}}, \
		and\ \bibinfo {author} {\bibfnamefont {A.}~\bibnamefont {Al\`u}},\ }\bibfield
	{title} {\enquote {\bibinfo {title} {Metagratings: Beyond the limits of
				graded metasurfaces for wave front control},}\ }\href {\doibase
		10.1103/PhysRevLett.119.067404} {\bibfield  {journal} {\bibinfo  {journal}
			{Phys. Rev. Lett.}\ }\textbf {\bibinfo {volume} {119}},\ \bibinfo {pages}
		{067404} (\bibinfo {year} {2017})}\BibitemShut {NoStop}%
	\bibitem [{\citenamefont {Epstein}\ and\ \citenamefont
		{Rabinovich}(2017)}]{Epstein2017}%
	\BibitemOpen
	\bibfield  {author} {\bibinfo {author} {\bibfnamefont {A.}~\bibnamefont
			{Epstein}}\ and\ \bibinfo {author} {\bibfnamefont {O.}~\bibnamefont
			{Rabinovich}},\ }\bibfield  {title} {\enquote {\bibinfo {title} {Unveiling
				the properties of metagratings via a detailed analytical model for synthesis
				and analysis},}\ }\href {\doibase 10.1103/PhysRevApplied.8.054037} {\bibfield
		{journal} {\bibinfo  {journal} {Phys. Rev. Applied}\ }\textbf {\bibinfo
			{volume} {8}},\ \bibinfo {pages} {054037} (\bibinfo {year}
		{2017})}\BibitemShut {NoStop}%
	\bibitem [{\citenamefont {Rabinovich}\ and\ \citenamefont
		{Epstein}(2018)}]{Epstein2018}%
	\BibitemOpen
	\bibfield  {author} {\bibinfo {author} {\bibfnamefont {O.}~\bibnamefont
			{Rabinovich}}\ and\ \bibinfo {author} {\bibfnamefont {A.}~\bibnamefont
			{Epstein}},\ }\bibfield  {title} {\enquote {\bibinfo {title} {Analytical
				design of printed circuit board (pcb) metagratings for perfect anomalous
				reflection},}\ }\href {\doibase 10.1109/TAP.2018.2836379} {\bibfield
		{journal} {\bibinfo  {journal} {IEEE Transactions on Antennas and
				Propagation}\ }\textbf {\bibinfo {volume} {66}},\ \bibinfo {pages}
		{4086--4095} (\bibinfo {year} {2018})}\BibitemShut {NoStop}%
	\bibitem [{\citenamefont {Epstein}\ and\ \citenamefont
		{Rabinovich}(2018)}]{epstein2018anrefr}%
	\BibitemOpen
	\bibfield  {author} {\bibinfo {author} {\bibfnamefont {A.}~\bibnamefont
			{Epstein}}\ and\ \bibinfo {author} {\bibfnamefont {O.}~\bibnamefont
			{Rabinovich}},\ }\bibfield  {title} {\enquote {\bibinfo {title} {Perfect
				anomalous refraction with metagratings},}\ }\href@noop {} {\bibfield
		{journal} {\bibinfo  {journal} {arXiv preprint arXiv:1804.02362}\ } (\bibinfo
		{year} {2018})}\BibitemShut {NoStop}%
	\bibitem [{\citenamefont {Rabinovich}\ \emph {et~al.}(2018)\citenamefont
		{Rabinovich}, \citenamefont {Kaplon}, \citenamefont {Reis},\ and\
		\citenamefont {Epstein}}]{rabinovich2018experimental}%
	\BibitemOpen
	\bibfield  {author} {\bibinfo {author} {\bibfnamefont {O.}~\bibnamefont
			{Rabinovich}}, \bibinfo {author} {\bibfnamefont {I.}~\bibnamefont {Kaplon}},
		\bibinfo {author} {\bibfnamefont {J.}~\bibnamefont {Reis}}, \ and\ \bibinfo
		{author} {\bibfnamefont {A.}~\bibnamefont {Epstein}},\ }\bibfield  {title}
	{\enquote {\bibinfo {title} {Experimental demonstration and in-depth
				investigation of analytically designed anomalous reflection metagratings},}\
	}\href@noop {} {\bibfield  {journal} {\bibinfo  {journal} {arXiv preprint
				arXiv:1809.01938}\ } (\bibinfo {year} {2018})}\BibitemShut {NoStop}%
	\bibitem [{\citenamefont {Wong}\ and\ \citenamefont
		{Eleftheriades}(2018)}]{Eleftheriades_2018}%
	\BibitemOpen
	\bibfield  {author} {\bibinfo {author} {\bibfnamefont {A.~M.~H.}\
			\bibnamefont {Wong}}\ and\ \bibinfo {author} {\bibfnamefont {G.~V.}\
			\bibnamefont {Eleftheriades}},\ }\bibfield  {title} {\enquote {\bibinfo
			{title} {Perfect anomalous reflection with a bipartite huygens'
				metasurface},}\ }\href {\doibase 10.1103/PhysRevX.8.011036} {\bibfield
		{journal} {\bibinfo  {journal} {Phys. Rev. X}\ }\textbf {\bibinfo {volume}
			{8}},\ \bibinfo {pages} {011036} (\bibinfo {year} {2018})}\BibitemShut
	{NoStop}%
	\bibitem [{\citenamefont {Wong}\ \emph {et~al.}(2018)\citenamefont {Wong},
		\citenamefont {Christian},\ and\ \citenamefont
		{Eleftheriades}}]{Eleftheriades_2018_1}%
	\BibitemOpen
	\bibfield  {author} {\bibinfo {author} {\bibfnamefont {A.~M.~H.}\
			\bibnamefont {Wong}}, \bibinfo {author} {\bibfnamefont {P.}~\bibnamefont
			{Christian}}, \ and\ \bibinfo {author} {\bibfnamefont {G.~V.}\ \bibnamefont
			{Eleftheriades}},\ }\bibfield  {title} {\enquote {\bibinfo {title} {Binary
				huygens’ metasurfaces: Experimental demonstration of simple and efficient
				near-grazing retroreflectors for te and tm polarizations},}\ }\href {\doibase
		10.1109/TAP.2018.2816792} {\bibfield  {journal} {\bibinfo  {journal} {IEEE
				Transactions on Antennas and Propagation}\ }\textbf {\bibinfo {volume}
			{66}},\ \bibinfo {pages} {2892--2903} (\bibinfo {year} {2018})}\BibitemShut
	{NoStop}%
	\bibitem [{\citenamefont {Popov}\ \emph {et~al.}(2018)\citenamefont {Popov},
		\citenamefont {Boust},\ and\ \citenamefont {Burokur}}]{Popov2018}%
	\BibitemOpen
	\bibfield  {author} {\bibinfo {author} {\bibfnamefont {V.}~\bibnamefont
			{Popov}}, \bibinfo {author} {\bibfnamefont {F.}~\bibnamefont {Boust}}, \ and\
		\bibinfo {author} {\bibfnamefont {S.~N.}\ \bibnamefont {Burokur}},\
	}\bibfield  {title} {\enquote {\bibinfo {title} {Controlling diffraction
				patterns with metagratings},}\ }\href {\doibase
		10.1103/PhysRevApplied.10.011002} {\bibfield  {journal} {\bibinfo  {journal}
			{Phys. Rev. Applied}\ }\textbf {\bibinfo {volume} {10}},\ \bibinfo {pages}
		{011002} (\bibinfo {year} {2018})}\BibitemShut {NoStop}%
	\bibitem [{\citenamefont {Tretyakov}(2003)}]{tretyakov2003analytical}%
	\BibitemOpen
	\bibfield  {author} {\bibinfo {author} {\bibfnamefont {S.}~\bibnamefont
			{Tretyakov}},\ }\href@noop {} {\emph {\bibinfo {title} {Analytical modeling
				in applied electromagnetics}}}\ (\bibinfo  {publisher} {Artech House},\
	\bibinfo {year} {2003})\BibitemShut {NoStop}%
	\bibitem [{\citenamefont {Luukkonen}\ \emph {et~al.}(2008)\citenamefont
		{Luukkonen}, \citenamefont {Simovski}, \citenamefont {Granet}, \citenamefont
		{Goussetis}, \citenamefont {Lioubtchenko}, \citenamefont {Raisanen},\ and\
		\citenamefont {Tretyakov}}]{Tretyakov_patches_2008}%
	\BibitemOpen
	\bibfield  {author} {\bibinfo {author} {\bibfnamefont {O.}~\bibnamefont
			{Luukkonen}}, \bibinfo {author} {\bibfnamefont {C.}~\bibnamefont {Simovski}},
		\bibinfo {author} {\bibfnamefont {G.}~\bibnamefont {Granet}}, \bibinfo
		{author} {\bibfnamefont {G.}~\bibnamefont {Goussetis}}, \bibinfo {author}
		{\bibfnamefont {D.}~\bibnamefont {Lioubtchenko}}, \bibinfo {author}
		{\bibfnamefont {A.~V.}\ \bibnamefont {Raisanen}}, \ and\ \bibinfo {author}
		{\bibfnamefont {S.~A.}\ \bibnamefont {Tretyakov}},\ }\bibfield  {title}
	{\enquote {\bibinfo {title} {Simple and accurate analytical model of planar
				grids and high-impedance surfaces comprising metal strips or patches},}\
	}\href {\doibase 10.1109/TAP.2008.923327} {\bibfield  {journal} {\bibinfo
			{journal} {IEEE Transactions on Antennas and Propagation}\ }\textbf {\bibinfo
			{volume} {56}},\ \bibinfo {pages} {1624--1632} (\bibinfo {year}
		{2008})}\BibitemShut {NoStop}%
	\bibitem [{\citenamefont {Wang}\ \emph
		{et~al.}(2018{\natexlab{a}})\citenamefont {Wang}, \citenamefont
		{D\'iaz-Rubio}, \citenamefont {Sneck}, \citenamefont {Alastalo},
		\citenamefont {M\"akel\"a}, \citenamefont {Ala-Laurinaho}, \citenamefont
		{Zheng}, \citenamefont {R\"ais\"anen},\ and\ \citenamefont
		{Tretyakov}}]{Tretyakov_MetaAnalyt_2018}%
	\BibitemOpen
	\bibfield  {author} {\bibinfo {author} {\bibfnamefont {X.~C.}\ \bibnamefont
			{Wang}}, \bibinfo {author} {\bibfnamefont {A.}~\bibnamefont {D\'iaz-Rubio}},
		\bibinfo {author} {\bibfnamefont {A.}~\bibnamefont {Sneck}}, \bibinfo
		{author} {\bibfnamefont {A.}~\bibnamefont {Alastalo}}, \bibinfo {author}
		{\bibfnamefont {T.}~\bibnamefont {M\"akel\"a}}, \bibinfo {author}
		{\bibfnamefont {J.}~\bibnamefont {Ala-Laurinaho}}, \bibinfo {author}
		{\bibfnamefont {J.~F.}\ \bibnamefont {Zheng}}, \bibinfo {author}
		{\bibfnamefont {A.~V.}\ \bibnamefont {R\"ais\"anen}}, \ and\ \bibinfo
		{author} {\bibfnamefont {S.~A.}\ \bibnamefont {Tretyakov}},\ }\bibfield
	{title} {\enquote {\bibinfo {title} {Systematic design of printable
				metasurfaces: Validation through reverse-offset printed millimeter-wave
				absorbers},}\ }\href {\doibase 10.1109/TAP.2017.2783324} {\bibfield
		{journal} {\bibinfo  {journal} {IEEE Transactions on Antennas and
				Propagation}\ }\textbf {\bibinfo {volume} {66}},\ \bibinfo {pages}
		{1340--1351} (\bibinfo {year} {2018}{\natexlab{a}})}\BibitemShut {NoStop}%
	\bibitem [{\citenamefont {Wang}\ \emph
		{et~al.}(2018{\natexlab{b}})\citenamefont {Wang}, \citenamefont
		{D\'iaz-Rubio}, \citenamefont {Asadchy},\ and\ \citenamefont
		{Tretyakov}}]{wang2018reciprocal}%
	\BibitemOpen
	\bibfield  {author} {\bibinfo {author} {\bibfnamefont {X.-C.}\ \bibnamefont
			{Wang}}, \bibinfo {author} {\bibfnamefont {A.}~\bibnamefont {D\'iaz-Rubio}},
		\bibinfo {author} {\bibfnamefont {V.~S.}\ \bibnamefont {Asadchy}}, \ and\
		\bibinfo {author} {\bibfnamefont {S.~A.}\ \bibnamefont {Tretyakov}},\
	}\bibfield  {title} {\enquote {\bibinfo {title} {Reciprocal angle-asymmetric
				absorbers: Concept and design},}\ }\href@noop {} {\bibfield  {journal}
		{\bibinfo  {journal} {arXiv preprint arXiv:1801.09397}\ } (\bibinfo {year}
		{2018}{\natexlab{b}})}\BibitemShut {NoStop}%
	\bibitem [{\citenamefont {Stratton}\ and\ \citenamefont
		{Chu}(1939)}]{Chu_Stratton1939}%
	\BibitemOpen
	\bibfield  {author} {\bibinfo {author} {\bibfnamefont {J.~A.}\ \bibnamefont
			{Stratton}}\ and\ \bibinfo {author} {\bibfnamefont {L.~J.}\ \bibnamefont
			{Chu}},\ }\bibfield  {title} {\enquote {\bibinfo {title} {Diffraction theory
				of electromagnetic waves},}\ }\href {\doibase 10.1103/PhysRev.56.99}
	{\bibfield  {journal} {\bibinfo  {journal} {Phys. Rev.}\ }\textbf {\bibinfo
			{volume} {56}},\ \bibinfo {pages} {99--107} (\bibinfo {year}
		{1939})}\BibitemShut {NoStop}%
	\bibitem [{\citenamefont {Stratton}(2007)}]{stratton2007electromagnetic}%
	\BibitemOpen
	\bibfield  {author} {\bibinfo {author} {\bibfnamefont {J.~A.}\ \bibnamefont
			{Stratton}},\ }\href@noop {} {\emph {\bibinfo {title} {Electromagnetic
				theory}}}\ (\bibinfo  {publisher} {John Wiley \& Sons, Hoboken, New Jersey},\
	\bibinfo {year} {2007})\BibitemShut {NoStop}%
	\bibitem [{\citenamefont {Fan}\ \emph {et~al.}(2018)\citenamefont {Fan},
		\citenamefont {Shcherbakov}, \citenamefont {Allen}, \citenamefont {Allen},
		\citenamefont {Wenner},\ and\ \citenamefont {Shvets}}]{fan2018perfect}%
	\BibitemOpen
	\bibfield  {author} {\bibinfo {author} {\bibfnamefont {Z.}~\bibnamefont
			{Fan}}, \bibinfo {author} {\bibfnamefont {M.~R.}\ \bibnamefont
			{Shcherbakov}}, \bibinfo {author} {\bibfnamefont {M.}~\bibnamefont {Allen}},
		\bibinfo {author} {\bibfnamefont {J.}~\bibnamefont {Allen}}, \bibinfo
		{author} {\bibfnamefont {B.}~\bibnamefont {Wenner}}, \ and\ \bibinfo {author}
		{\bibfnamefont {G.}~\bibnamefont {Shvets}},\ }\bibfield  {title} {\enquote
		{\bibinfo {title} {Perfect diffraction with multiresonant bianisotropic
				metagratings},}\ }\href@noop {} {\bibfield  {journal} {\bibinfo  {journal}
			{ACS Photonics}\ }\textbf {\bibinfo {volume} {5}},\ \bibinfo {pages}
		{4303--4311} (\bibinfo {year} {2018})}\BibitemShut {NoStop}%
	\bibitem [{\citenamefont {{Popov}}\ \emph {et~al.}(2018)\citenamefont
		{{Popov}}, \citenamefont {{Yakovleva}}, \citenamefont {{Boust}},\ and\
		\citenamefont {{Burokur}}}]{2018arXiv181210164P}%
	\BibitemOpen
	\bibfield  {author} {\bibinfo {author} {\bibfnamefont {V.}~\bibnamefont
			{{Popov}}}, \bibinfo {author} {\bibfnamefont {M.}~\bibnamefont
			{{Yakovleva}}}, \bibinfo {author} {\bibfnamefont {F.}~\bibnamefont
			{{Boust}}}, \ and\ \bibinfo {author} {\bibfnamefont {S.~N.}\ \bibnamefont
			{{Burokur}}},\ }\bibfield  {title} {\enquote {\bibinfo {title} {{Designing
					Metagratings Via Local Periodic Approximation: From Microwaves to
					Infrared}},}\ }\href@noop {} {\bibfield  {journal} {\bibinfo  {journal}
			{arXiv e-prints}\ ,\ \bibinfo {eid} {arXiv:1812.10164}} (\bibinfo {year}
		{2018})},\ \Eprint {http://arxiv.org/abs/1812.10164} {arXiv:1812.10164
		[physics.app-ph]} \BibitemShut {NoStop}%
	\bibitem [{\citenamefont {Felsen}\ and\ \citenamefont
		{Marcuvitz}(1994)}]{felsen1994radiation}%
	\BibitemOpen
	\bibfield  {author} {\bibinfo {author} {\bibfnamefont {L.~B.}\ \bibnamefont
			{Felsen}}\ and\ \bibinfo {author} {\bibfnamefont {N.}~\bibnamefont
			{Marcuvitz}},\ }\href@noop {} {\emph {\bibinfo {title} {Radiation and
				scattering of waves}}},\ Vol.~\bibinfo {volume} {31}\ (\bibinfo  {publisher}
	{John Wiley \& Sons},\ \bibinfo {year} {1994})\BibitemShut {NoStop}%
	\bibitem [{\citenamefont {Li}\ \emph {et~al.}(2014)\citenamefont {Li},
		\citenamefont {Jiang}, \citenamefont {Li}, \citenamefont {Liang},
		\citenamefont {Zou}, \citenamefont {Yin},\ and\ \citenamefont
		{Cheng}}]{PhysRevApplied.2.064002}%
	\BibitemOpen
	\bibfield  {author} {\bibinfo {author} {\bibfnamefont {Y.}~\bibnamefont
			{Li}}, \bibinfo {author} {\bibfnamefont {X.}~\bibnamefont {Jiang}}, \bibinfo
		{author} {\bibfnamefont {R.}~\bibnamefont {Li}}, \bibinfo {author}
		{\bibfnamefont {B.}~\bibnamefont {Liang}}, \bibinfo {author} {\bibfnamefont
			{X.}~\bibnamefont {Zou}}, \bibinfo {author} {\bibfnamefont {L.}~\bibnamefont
			{Yin}}, \ and\ \bibinfo {author} {\bibfnamefont {J.}~\bibnamefont {Cheng}},\
	}\bibfield  {title} {\enquote {\bibinfo {title} {Experimental realization of
				full control of reflected waves with subwavelength acoustic metasurfaces},}\
	}\href {\doibase 10.1103/PhysRevApplied.2.064002} {\bibfield  {journal}
		{\bibinfo  {journal} {Phys. Rev. Applied}\ }\textbf {\bibinfo {volume} {2}},\
		\bibinfo {pages} {064002} (\bibinfo {year} {2014})}\BibitemShut {NoStop}%
	\bibitem [{\citenamefont {Li}\ \emph {et~al.}(2018)\citenamefont {Li},
		\citenamefont {Shen}, \citenamefont {D{\'\i}az-Rubio}, \citenamefont
		{Tretyakov},\ and\ \citenamefont {Cummer}}]{Diaz2018_acoustic}%
	\BibitemOpen
	\bibfield  {author} {\bibinfo {author} {\bibfnamefont {J.}~\bibnamefont
			{Li}}, \bibinfo {author} {\bibfnamefont {C.}~\bibnamefont {Shen}}, \bibinfo
		{author} {\bibfnamefont {A.}~\bibnamefont {D{\'\i}az-Rubio}}, \bibinfo
		{author} {\bibfnamefont {S.~A.}\ \bibnamefont {Tretyakov}}, \ and\ \bibinfo
		{author} {\bibfnamefont {S.~A.}\ \bibnamefont {Cummer}},\ }\bibfield  {title}
	{\enquote {\bibinfo {title} {Systematic design and experimental demonstration
				of bianisotropic metasurfaces for scattering-free manipulation of acoustic
				wavefronts},}\ }\href@noop {} {\bibfield  {journal} {\bibinfo  {journal}
			{Nature communications}\ }\textbf {\bibinfo {volume} {9}},\ \bibinfo {pages}
		{1342} (\bibinfo {year} {2018})}\BibitemShut {NoStop}%
	\bibitem [{\citenamefont {Torrent}(2018)}]{PhysRevB.98.060101}%
	\BibitemOpen
	\bibfield  {author} {\bibinfo {author} {\bibfnamefont {D.}~\bibnamefont
			{Torrent}},\ }\bibfield  {title} {\enquote {\bibinfo {title} {Acoustic
				anomalous reflectors based on diffraction grating engineering},}\ }\href
	{\doibase 10.1103/PhysRevB.98.060101} {\bibfield  {journal} {\bibinfo
			{journal} {Phys. Rev. B}\ }\textbf {\bibinfo {volume} {98}},\ \bibinfo
		{pages} {060101} (\bibinfo {year} {2018})}\BibitemShut {NoStop}%
	\bibitem [{\citenamefont {Packo}\ \emph {et~al.}(2019)\citenamefont {Packo},
		\citenamefont {Norris},\ and\ \citenamefont
		{Torrent}}]{PhysRevApplied.11.014023}%
	\BibitemOpen
	\bibfield  {author} {\bibinfo {author} {\bibfnamefont {P.}~\bibnamefont
			{Packo}}, \bibinfo {author} {\bibfnamefont {A.~N.}\ \bibnamefont {Norris}}, \
		and\ \bibinfo {author} {\bibfnamefont {D.}~\bibnamefont {Torrent}},\
	}\bibfield  {title} {\enquote {\bibinfo {title} {Inverse grating problem:
				Efficient design of anomalous flexural wave reflectors and refractors},}\
	}\href {\doibase 10.1103/PhysRevApplied.11.014023} {\bibfield  {journal}
		{\bibinfo  {journal} {Phys. Rev. Applied}\ }\textbf {\bibinfo {volume}
			{11}},\ \bibinfo {pages} {014023} (\bibinfo {year} {2019})}\BibitemShut
	{NoStop}%
\end{thebibliography}

%

\end{document}